\documentclass{aa}  
\pdfoutput=1
\usepackage{verbatim}
\usepackage{subfigure}
\usepackage{graphicx}
\usepackage{caption}
\usepackage{printlen}
\usepackage{amsfonts} 
\usepackage{amssymb}
\usepackage{amsmath}
\usepackage{latexsym}
\usepackage[T1]{fontenc} 
\usepackage{footnote}
\usepackage{graphicx}
\usepackage{txfonts}
\usepackage{hyperref}
\usepackage{ulem}
\usepackage{color}

\usepackage{natbib}
\bibpunct{(}{)}{;}{a}{}{,} 
\hypersetup{%
  colorlinks=true,
  citecolor=cyan
}

\def\lm{{\ell m}}
\def\inv{^{-1}}

\begin{document}

   \title{The CMB angular power spectrum via component separation: \\ 
   a study on Planck data}

\author{C. Umilt\`a \inst{1} \inst{2} \inst{3}  \and J.F. Cardoso \inst{1} \and K. Benabed \inst{1} \and M. Le Jeune \inst{4}}

\institute{Institut d'Astrophysique de Paris, Sorbonne Universit\'e, CNRS (UMR7095), 98 bis Boulevard
Arago, F-75014, Paris, France
\and
Sorbonne Universit\'es, Institut Lagrange de Paris (ILP), 98 bis Boulevard Arago, 75014 Paris, France
\and
University of Cincinnati, Cincinnati, Ohio 45221
\and
APC, Univ Paris Diderot, CNRS/IN2P3, CEA/Irfu, Obs de Paris, Sorbonne Paris Cité, France}

  \abstract
{}
  { We investigate the extent to which foreground cleaned CMB maps can be used to 
  estimate the cosmological parameters at small scales.}
  { We use the SMICA method,  a blind separation technique which works directly
    at the spectral level. In this work we focus on the small scales
    of the CMB angular power spectrum, which are chiefly affected by
    noise and extragalactic foregrounds, such as point sources. We
    adapt SMICA to use only cross-spectra between data maps, thus
    avoiding the noise bias. In this study, performed both on simulations and on Planck 2015 data, 
    we fit for extragalactic point sources by modeling them as shot noise of two independent populations. }
  { In simulations we correctly recover the point source emission law,
    and obtain a CMB angular power spectrum that has an average
    foreground residual of one fifth of the CMB power at
    $\ell \geq$2200. On Planck data, the recovered point source
    emission law corresponds to external estimates, with some offsets
    at the highest and lowest frequencies, possibly due to frequency
    decoherence of point sources. The CMB angular power spectrum
    residuals are consistent with what we find in simulations. The
    cosmological parameters obtained from the simulations and the data
    show offsets up to 1$\sigma$ on average from their expected
    values. Biases on cosmological parameters in simulations represent
    the expected level of bias in Planck data. } 
    {The results on cosmological parameters depend on the detail of the foreground
    residual contamination in the spectrum, and need a tailored
    modeling of the likelihood foreground model.}
   
   
   


   \maketitle

\section{Introduction}
The cosmic microwave background (CMB) is an important probe for cosmology, and in recent
years a lot of effort has been dedicated to its extraction from available data. In
particular, the CMB angular power spectrum can be used to constrain the cosmological parameters. 
The Planck mission provided the astronomical community
with full-sky observations in nine frequency bands between 30 and 857 GHz. These data are 
extremely important for CMB science since they allow to characterize and separate the primordial CMB signal from the other astrophysical galactic and extragalactic emissions or \textit{foregrounds}.\\
Separating the CMB from the foregrounds is a highly non trivial task, and a
number of component separation methods has been conceived in the past years.
The Commander method  \citep{Eriksen:2004,Eriksen:2008} performs a Bayesian exploration of a physical parametric model.
The SEVEM method \citep{Fernandez:2012} fits a number of templates obtained from the data
themselves, while the  CCA \citep{Bedini:2005,Bonaldi:2006} technique exploits the foregrounds spatial correlations to recover their emission law.\\
Blind techniques, that use no prior information on foreground emission, have been largely
employed. The advantage of such an approach is that it does not need any assumption on the foreground
contamination.  We note that this is also the case for SEVEM. Among blind methods, NILC \citep{Delabrouille:2009,Basak:2012,Basak:2013}, GNILC
\citep{Remazeilles:2011b} and more recently SILC \citep{Rogers:2016} and HILC \citep{Sudevan_2016} all exploit the ILC
technique. The L-GMCA \citep{Bobin:2008, Bobin:2013} method exploits the sparsity of the data in the wavelet domain. \\
Another popular technique for blind source separation is the Independent Component
Analysis (ICA), implemented by FastICA \citep{Maino:2002}, which is based on the
non-Gaussianity of the sources, BICA \citep{Vansyngel:2014} which implements ICA in a
Bayesian framework and SMICA \citep{Cardoso:2008}, which blindly recovers the sources spectra in the maximum likelihood sense. 
The effort in finding new tools for component separation is still on-going: for example the ABS blind method has been recently proposed \citep{Zhang:2016}. \\
The {\it Planck Collaboration} has selected four different methods to perform component
separation and produce CMB maps.  These are Commander, NILC, SEVEM and SMICA \citep{P15_compsep}. Nevertheless, 
the cosmological analysis of the CMB is very sensitive to residual foreground
contamination in the data. The angular power spectra derived from the four Planck CMB maps 
have a residual foreground content which prevents their use for cosmological purposes. 
In particular, the small scale residuals of unresolved point sources in these maps is not well characterized \citep{P15_lkl}.
Among the four CMB maps released by the {\it Planck Collaboration}, the SMICA map has the lowest
extragalactic contribution at high-$\ell$ \citep{P15_compsep}.  
More importantly for this work, SMICA proceeds in two steps: a component separation at the
spectral (harmonic) level is first performed, the result of which is then used to control
the synthesis of the CMB map from its harmonic coefficients.
It is the first step of SMICA ---component separation at the spectral level--- which is of
interest in this contribution.\\
In this work, we shall consider variations on the SMICA method of component separation,
targeting direct cosmological analysis, based on the following three ideas.
First, we take advantage of the `data splits' available in Planck products.  
By cross-correlating the two halves of a data split, one obtains spectral estimates which
are free of noise bias \citep{Hinshaw:2003}, at the cost of a reasonable variance increase.  \\
Second, we build on the fact that the SMICA method is based on a statistical fit
\textit{in the spectral domain} of a model of independent spectra (CMB, foregrounds and
noise), the cleaned-up CMB map being produced in a second (optional) step.  In this work,
we shall focus on the first step of spectral fitting, \textit{i.e.},  the SMICA method is
only used as a tool for the joint fit of all the auto- and cross-spectra of a set of
frequency channels. In this way the CMB angular power spectrum is estimated directly from the data.\\
Third, we introduce constraints on some of the spectral components fitted by SMICA.  
In the standard operation of SMICA, the foreground contribution is fully unconstrained \citep{P13_lkl}.
This design choice naturally targets the large scale galactic contamination, failing to
remove accurately the extragalactic contamination, since the latter is subdominant with
respect to the galactic one on a wide range of scales. In this work, we use a foreground model 
that targets the extragalactic emission of unresolved point sources. In particular, we model them as two
independent populations with a shot noise angular power spectrum. This allows us to recover 
their emission law, at the price of a partial loss of blindness
of the method. \\
The natural comparison of this work results is with the cosmological analysis of the {\it Planck Collaboration} \citep{P15_lkl, P15_cosmopar} .
Their high-$\ell$ likelihood (PlikTT) is based on the spectra of a few frequency channels with low foreground content, 
in the cleaner area of the sky and on
a tailored scale range.  In this likelihood, the residual foreground contamination is
described by a set of templates controlled by a few parameters for each non-negligible
astrophysical contribution. The extragalactic point sources are modeled with a free amplitude parameter at each frequency. \\
This paper is organised as follows. 
In Sect.~\ref{sec:foreground}, we introduce the foreground emissions relevant to this analysis.
In Sect.~\ref{sec:new_method}, we present the SMICA method and how we adapt it to spectral component separation based on cross-spectra.
In Sect.~\ref{sec:data} we present the data used and in Sect.~\ref{sec:test}
we present the results of the SMICA fit.
Finally we show cosmological parameters obtained from the fitted CMB angular power spectra in Sect.~\ref{sec:cosmo}.
Our conclusions are given in Sect.~\ref{sec:conclusion}.
\section{The astrophysical foregrounds}\label{sec:foreground}
The Planck and WMAP satellites delivered a set of full sky maps in the 23 $\leq \nu \leq$ 857 GHz range. Component separation methods aimed at map reconstruction exploit a wide range
of frequencies in order to ameliorate their cleaning efficiency close to the galactic center.
In this work we want to obtain a CMB power spectrum with low foreground contamination in particular at small scales and with respect to extragalactic contamination.
Even though we are interested in retaining the largest possible sky fraction, the complexity of foreground contamination close to the galactic 
center is a considerable drawback and we prefer to exclude this region from the analysis. We limit ourselves  to frequencies larger than 100 GHz, where lower frequency galactic contamination from free-free and synchrotron is negligible \citep{P15_diff_low}.
We present here a brief description of the foreground contamination at the frequencies of interest of this work.
The relevant foreground emissions at frequencies $\nu \geq$ 100 GHz are the thermal dust emission from our galaxy, 
and the emission of background unresolved galaxies.
\subsection{Thermal dust}
The galactic dust is the dominant foreground at large scales for frequencies above 70 GHz \citep{2014PTEP.2014fB109I}. 
Its emission law can be empirically described by a modified black-body: 
\begin{equation}
 I_{\nu} \propto \nu^{\beta_d} B_{\nu}(T), 
\end{equation}\label{dust_em}
where $B_{\nu}(T)$ is the Planck black-body spectrum at a temperature $T$, while $\beta_d$ is the dust spectral index. 
The values of $T$ and $\beta_d$ vary across the sky: as a reference, we can consider a temperature  $T=(19.4\pm 1.3) $ K and 
an average spectral index $\beta_d = 1.6\pm 0.1$ \citep[Sect. 4.2.2]{P15_gnilc}.\\
We can describe the angular power spectrum of dust approximately  as \citep[Sect. 3.3]{P15_lkl}:
\begin{equation}\label{dust_cl}
 C_{\ell}^{dust} \propto \ell^{-2.6},
\end{equation}\label{dust_cl}
thus the dust contribution drops quickly at small scales. 
\subsection{Extragalactic foregrounds}\label{sec:extrafg}
The extragalactic contamination, which is only relevant at small angular scales, comes essentially from background galaxies and clusters. 
The former produce the radio point sources and Cosmic Infrared Background (CIB) contamination, while the latter 
are responsible for the SZ effect. 
The SZ effect, which is the distortion on CMB photons produced by the interaction with the intra-cluster hot gas, 
is not well constrained by Planck data alone \citep{P15_lkl}.  
In the Planck likelihood analysis, this emission has been constrained by using SPT
and ACT small scale data \citep{P13_lkl} or by imposing a narrow
prior \citep{P15_lkl}. 
The background galaxies instead are an important source of contamination in Planck data. The resolved sources are masked, 
but a background diffuse emission of unresolved point sources is
still present in the maps. \\
These sources can be categorized in two types: red elliptical galaxies that emit essentially in the radio band and
dusty star-forming galaxies, which are more luminous in the infrared (IR) and produce the CIB emission. 
The point source emission can be described by at least two contributions, a shot noise part, due to the average random distribution of galaxies, 
and a clustered part due to the fact that galaxies follow the matter distribution and are thus not evenly distributed on the sky. 
Contrary to the radio sources, which are  well described by the shot noise model alone \citep{Hall:2010, Lacasa:2012}, the infrared clustered contribution is quite important, and we refer to it as clustered CIB.
The shot noise contribution has a flat spectrum \citep{Tegmark1996}:
\begin{equation}
 C_{\ell}^{shot}= C^{shot}.
\end{equation}
The clustered CIB contribution can be represented by:
\begin{equation}
C_{\ell}^{clustered}= \ell^{\alpha},
\end{equation}
where $\alpha$ = -1.4 for $\ell$ > 2500, and is shallower at larger scales  \citep[Sect. 3.3]{P15_lkl}.

\section{Spectral component separation}
\label{sec:new_method}
\label{sec:smica}
The SMICA method \citep{Cardoso:2008} ({\it Spectral Matching Independent Component
  Analysis}) is a blind component separation method that works at the spectral
level. SMICA is one of the four component separation tools used by the {\it Planck  Collaboration} for map reconstruction. SMICA works by adjusting a model $R_\ell(\theta)$ to a set of `spectral covariance matrices' $\hat{R}_{\ell}$ derived from the data. These matrices are defined as follows.\\
Given a set of $n$ observed sky maps in $n$ frequency channels, we denote $y_{\ell m}^i$\footnote{We use real valued spherical harmonics.} 
their spherical harmonic coefficients ($i=1,\ldots,n$) and we denote $\bold{y}_{\ell m}$ the
$n\times1$ vector which collects them.  These observations are made of signal $\bold{o}_{\ell m}$ and noise $\bold{n}_\lm$:
\begin{equation}
  \bold{y}_{\ell m} = \bold{o}_{\ell m} + \bold{n}_\lm.
\end{equation}
Using these $ \bold{y}_{\ell m}$ coefficients, the auto- and cross spectra of the input maps can be computed and collected in $n\times n$ empirical spectral covariance matrices $\hat{R}_{\ell}$ defined as:
\begin{equation}\label{smi_covmat}
 \hat{R}_{\ell}= \frac{1}{2\ell+1}\sum_m \bold{y}_{\ell m}\ \ \bold{y}_{\ell m}^{T}.
\end{equation}
These matrices contain at each angular frequency $\ell$ the auto-spectra of each channel in their
diagonal entries and the respective cross-spectra in their off-diagonal entries.\\
The model $R_\ell(\theta)$ describes the expected value of the spectra in $\hat R_\ell$, and it has a tunable level of blindness, depending on the problem at hand. We will later see its specifications for the present work. The parameters of the model are adjusted to the data through the spectral matching criterion described in the next paragraph. 

\subsection{Spectral matching criterion}\label{SMICA:specmatchcrit}
The spectral fitting criterion used by SMICA is the likelihood obtained by assuming that
all input sky maps jointly follow a Gaussian stationary distribution.  
By standard arguments, one finds that for full sky statistics, the joint likelihood
depends only on $R_\ell(\theta)$ and is proportional to $\exp \Bigl(- \phi(\theta) \Bigr)$ where:
\begin{equation}
  \phi(\theta) = \frac12 \sum_\ell (2\ell+1) 
  \left[
    \mathrm{trace}(\hat{R}_\ell R_\ell(\theta)\inv)+\log\det R_\ell(\theta)
  \right]
  \ +\ \mathrm{cst}
  .
\end{equation}
It is useful to notice that, up to a constant term, this is equal to :
\begin{equation}
  \phi(\theta) =  \sum_\ell (2\ell+1) 
  K(\hat{R}_\ell, R_\ell(\theta))
  \ +\ \mathrm{cst'},
\end{equation}\label{mismatch}
where $K (R_1,R_2)$ is the Kullback-Leibler divergence defined as:
\begin{equation}
  K (R_1,R_2) =
  \frac{1}{2} \bigl[ \mathrm{trace} ( R_1 R_2\inv) - \log \det (R_1  R_2\inv) - n  \bigr],
\end{equation}
which measures the `divergence' between two $n\times n$ positive matrices $R_1$
and $R_2$.

\subsection{A new SMICA configuration}
In its regular mode of operation (as used in the Planck analysis, for instance), the SMICA
method has two main ingredients: a semi-blind model $R_\ell(\theta)$ for the expected
value of the spectra in $\hat R_\ell$ and a fitting criterion quantifying the discrepancy between $\hat R_\ell$ and $R_\ell(\theta)$.  
As discussed next, both these ingredients need to be adjusted for the present work, which addresses small scales limitations of the CMB angular power spectrum estimation, such as noise and point sources. 

\subsubsection{Data splits}\label{sec:data-splits}

The angular power spectra of sky maps always contain a noise term which needs to be
accurately characterized in order to avoid bias, especially at small scales.
Characterization of noise is often not trivial.  Noise derives from the instrumental
measurement and processing chain, making its
properties difficult to establish.\\
One possibility is to compare the noise contribution in data splits. These splits are
obtained by dividing in two halves the time-ordered data sequences.
For sky maps, this consists in generating the map with just half of the time ordered
information.  Therefore each data split contains the same astrophysical signal, but has a
different noise contribution.\\
In practice this means that the observations leading to $\bold{y}_{\ell m}$ are
split in two parts and processed independently, yielding two sky maps and therefore two
sets $\bold{y}_{\ell m}^a$ and $\bold{y}_{\ell m}^b$ of harmonic coefficients such that:
\begin{equation}
  \bold{y}_{\ell m}^a = \bold{o}_{\ell m} + \bold{n}_\lm^a
  \quad\text{and}\quad
  \bold{y}_{\ell m}^b = \bold{o}_{\ell m} + \bold{n}_\lm^b,
\end{equation}
where the noise coefficients $\bold{n}_\lm^{a}$ and $\bold{n}_\lm^{b}$ are assumed to
be uncorrelated.  
In the typical and simplest case of a balanced data split, one has:
\begin{equation}
  \bold{y}_{\ell m} 
  = \frac12 \left( \bold{y}_{\ell m}^a + \bold{y}_{\ell m}^b\right) 
  = \bold{o}_\lm + \frac12 \left( \bold{n}_{\ell m}^a + \bold{n}_{\ell m}^b\right) 
  = \bold{o}_\lm + \bold{n}_{\ell m}.
\end{equation}
The regular SMICA method uses as input spectral covariance matrices defined as in Eq. (\ref{smi_covmat}).
In this work, we consider using instead special matrices defined by:
\begin{equation}\label{eq_Rsplit}
 \hat{R}_{\ell}^{split}
 =
 \frac{1}{2\ell+1}
 \sum_m
 \frac12 
 \left(
   \bold{y}_\lm^a  \bold{y}_\lm^b{}^T +
   \bold{y}_\lm^b  \bold{y}_\lm^a{}^T 
 \right)
 ,
\end{equation}
 where the sum of the two terms is necessary in order to symmetrize the matrix. On average $\hat{R}_{\ell}^{split}$ represents correctly the sky, but we need to take into account its statistical properties. The first term of Eq.~(\ref{eq_Rsplit}) can be expanded as:
\begin{equation}\label{eq_R_split_singleterm}
\begin{split}
  \frac{1}{2\ell+1}
 \sum_m
  \bold{y}_\lm^a  \bold{y}_\lm^b{}^T
 =
  \frac{1}{2\ell+1}
 \sum_m
  (
  &\bold{o}_\lm^a  \bold{o}_\lm^b{}^T +
 \bold{o}_\lm^a  \bold{n}_\lm^b{}^T  +\\
+  &\bold{n}_\lm^a  \bold{o}_\lm^b{}^T+
  \bold{n}_\lm^a  \bold{n}_\lm^b{}^T ),
\end{split}
\end{equation}
By construction, these matrices contain only correlations between maps with independent
noise realizations and therefore they have a zero-mean noise contribution. 
More specifically, if we denote $\langle\cdot\rangle_N$ the average over noise
realisations, one has: 
\begin{equation}
  \langle  \hat{R}_{\ell} \rangle_N = \widehat{O}_\ell + N_\ell
  \qquad\text{but}\qquad
  \langle  \hat{R}_{\ell}^{split} \rangle_N = \widehat{O}_\ell, 
\end{equation}
where the sky part contribution (not averaged over) is:
\begin{equation}
  \widehat{O}_\ell   = \frac{1}{2\ell+1}\sum_m \bold{o}_{\ell m}\ \ \bold{o}_{\ell m}^{T},
\end{equation}
and where $N_\ell$ is the diagonal matrix with the noise spectra on its diagonal.
 The last three terms of Eq. (\ref{eq_R_split_singleterm}) are zero on average, but not for a single realisation. In practice, there are chance correlations between the CMB and noise,
which contribute to the scatter of the $\hat{R}_{\ell}^{split}$ matrix. 
So we need to take into account the fact
that for a single realisation of the data, matrix $\hat{R}_{\ell}^{split}$ is \emph{not}
distributed as $\hat{R}_{\ell}$ or even as $\hat{R}_{\ell} - N_\ell$.  
How the noise-unbiased spectra contained in matrices $\hat{R}_{\ell}^{split}$ are
jointly statistically fitted in the SMICA approach is described in next section.

\subsubsection{Spectral matching criterion using data splits}\label{SMICA:specmatchcrit}

We now consider using the SMICA criterion with the noise-unbiased spectral statistics
$\hat R_\ell^\mathrm{split}$.  
Let us denote $O_\ell(\theta)$ the expected value of $\hat R_\ell^\mathrm{split}$
since this is also the expected value of $\widehat{O}_\ell$.
It would be naive to adjust the spectral model $O_\ell(\theta)$ by minimizing
$\phi(\theta) = \sum_\ell (2\ell+1) K(\hat R_\ell^\mathrm{split}, O_\ell(\theta))$.
To see that, consider the divergence between two matrices which are close to each
other.  The second order (quadratic) approximation of the divergence is:
\begin{equation}\label{eq:quadapprox}
  K (R, R+\delta R) \approx
  K (R+\delta R, R) \approx
  \mathrm{trace}
  \bigl( 
  \delta R\, R\inv \, \delta R\, R\inv
  \bigr) / 4,
\end{equation}
and it shows that the Gaussian likelihood penalizes the (small) deviations $\delta R$
between covariance matrices through the inverse matrix $R\inv$.  This is the proper weight
(according to the maximum likelihood principle) to take into account the statistical
variability in sample covariance matrices.
Hence, if we were to use $K(\hat R_\ell^\mathrm{split}, O_\ell(\theta))$, the
statistical weight would not take into account the variability due to presence of the
noise in the spectra.  In order to account for that variability, we use an ansatz and
minimize:
\begin{equation}
  \phi^\mathrm{split}(\theta) =  \sum_\ell (2\ell+1) 
  K(\hat{R}_\ell^\mathrm{split} + N_\ell^\mathrm{eff} ,\ O_\ell(\theta) + N_\ell^\mathrm{eff}),
\end{equation}
where $N_\ell^\mathrm{eff}$ is a deterministic diagonal matrix containing the noise spectra which represents the effective noise contribution.  Since it is
introduced additively in both arguments of the $K(\cdot,\cdot)$, it should not introduce
noise bias (see Eq.~(\ref{eq:quadapprox})).\\
In the standard SMICA configuration, the minimum $\phi(\theta)$ is of the order of the number of degrees of freedom $d$ in the fit. We term this quantity 
`mismatch'. The mismatch is a diagnostic of the fit: high values indicate poor convergence or poor modeling.
In this work we present a SMICA configuration based on data splits only, 
in which the statistical properties of the covariance matrices are only approximately represented by the model. 
In this case the recovered mismatch $ \phi^\mathrm{split}(\theta)$ for a converged fit is not $\sim d$, and its value is difficult to 
predict\footnote{Nevertheless, for our configuration, simulations show that it is of the same order of magnitude of $d$.}. 
Even though we do not have a predicted value, 
the mismatch is still an interesting quantity to look at, since very high values indicate that the model cannot represent the data complexity.

\subsubsection{Semi-blind model}

In the standard use of SMICA, a fully non-parametric model is used to model the foreground
emission.  We postulate that the sky emission can be represented in the harmonic domain by:
\begin{equation}
 \bold{o}_{\ell m}= A\bold{s}_{\ell m},
\end{equation}
where $A$ is a fixed (independent of $\ell$) matrix of size $n\times(k+1)$
Its first column, denoted $\bold{a}$, is the spectral  energy  distribution (SED) of the CMB and the first entry of vector
$\bold{s}_\lm$ contains the CMB harmonic coefficients.
The remaining $k$ columns of matrix $A$ (resp. the last $k$ entries of $\bold{s}_\lm$)
represent foreground emissions.
Since the CMB is statically independent from the foregrounds, one has:
\begin{equation}\label{eq_model}
  O_\ell
  = 
  \begin{bmatrix}
    \bold{a} & F \\
  \end{bmatrix}
  \
  \begin{bmatrix}
    C_\ell^\mathrm{cmb} & 0 \\
    0 & P_\ell 
  \end{bmatrix}
  \
  \begin{bmatrix}
    \bold{a} & F \\
  \end{bmatrix}^T,
\end{equation}
where the $k\times k$ matrix $P_\ell$ is the covariance matrix of the foregrounds. 
The sources are modelled as Gaussian isotropic signals, since the only information retained
is their angular power spectrum and their emission law in frequency. 
Even though this approximation does not hold for foregrounds, 
it does not affect CMB recovery as long as the emission law of CMB is well known \citep{Cardoso2017}.\\
In the usual SMICA model, the $n\times k$ matrix $F$ is unconstrained and the symmetric
matrix $P_\ell$ is only constrained to be non-negative.  This amounts to saying that
foreground emission can be represented by $k$ templates with arbitrary SEDs, arbitrary
angular spectra and arbitrary correlation.\\
In this work, we shall consider a more constrained foreground model.  Those constraints
include forcing zero-terms in matrix $P_\ell$ (for instance to express independence
between point source emission and Galactic emission) as well as imposing a spectral
dependence to some entries (for instance flat angular spectra for point sources).
This is discussed and detailed in next section.

\subsection{Parametric models of foreground emission}\label{sec:smica3}

A strength of the regular SMICA approach is that very little assumptions are made
regarding the foreground emissions.  In the case of Planck data analysis, it is safe to assume
that the CMB has a black body emission law, within calibration errors.  
Nonetheless, nothing can be said about the other parameters: in an implementation as in Eq. (\ref{eq_model}), the foregrounds are described as a multidimensional component whose spectrum and emission law are totally free.\\
Some of the foreground emissions, in particular galactic emissions, present a degree of correlation 
that prevents their description as separate components.
If two emissions are not independent, then ICA methods, on which SMICA is based, can not separate them. 
Thus all dependent emissions must be grouped in the analysis and considered as one single multidimensional component \citep{Cardoso:2008}.
In the case of the model in Eq. (\ref{eq_model}), all the foreground emissions are grouped in one single component $P_\ell$. 
The existence of a correspondence between the spectra of this matrix 
with a given physical foreground emission is not guaranteed.\\
Large scale foregrounds, such as thermal dust, are well fit by multidimensional foregrounds, since the extra dimensions
account for the spatial variability of the foreground emission and its eventual correlations with other foregrounds.
Instead, small scale foregrounds, such as point sources, dominate in a region of the angular power spectrum where noise becomes important, are are less favoured by the fit. When using a large multidimensional component these foregrounds may not be accounted for correctly.
While this is less of an issue for the map reconstruction with SMICA, which aims at performing well at large scales \citep{P15_nongauss,P15_lensing},
it can be a serious drawback when using the fit results for cosmological estimation,  since separating the unresolved small scale foregrounds and the CMB power spectrum is difficult due to noise.\\ 
For this reason, as it is done in \citet{Patanchon:2005,P15_lkl}, we parameterize the foreground model. In particular, in this work we use a semi-blind model, by enforcing some minimal constraints on the extragalactic contamination. \\
The main sources of foreground contamination at the frequencies of interest of this study
are thermal dust, extragalactic shot noise and clustered contamination from point
sources. As detailed in Sect. \ref{sec:foreground}, the shot noise point sources emission
can be divided into a radio and an infrared component. The clustered point source
contamination instead, only
originates from infrared galaxies.  We build the foreground model as the sum of three
uncorrelated components: a bidimensional component that accounts for dust and clustered CIB (cCIB), and two unidimensional components to account for unresolved radio and infrared point
sources:
\begin{equation}\label{SMICA:model}
  \def\zero{\ \ \, 0 \ \ \,}
  \def\zeromat{\begin{array}{cc} \zero & \zero \\ \zero & \zero \end{array}}
  \def\psrcmat{\begin{array}{cc} P^{rad}_{\ell} &  0   \\  0  & P^{ir}_{\ell} \end{array}}
  P_{\ell} = 
  \left[
    \begin{array}{c@{}c}
      P^{dust + cCIB}_{\ell}   & \zeromat \\
      \zeromat & \psrcmat
    \end{array}
  \right].
\end{equation}
In practice, the clustered CIB is fitted together with thermal dust since they have an emission law which is very similar, and thus it is difficult to blindly identify them \citep{Delabrouille:2002kz}.\\ 
This model does not account for all the foreground contamination. In particular, both cCIB and dust present spatial variations in their spectral properties and would require more spectral dimensions. For point sources, we assume perfect coherence in frequency, which may not be true, but this would also require to describe them as multidimensional components (see for example \citet{marius,Paoletti:2012} for similar parameterizations of the extragalactic foregrounds). The dimensionality of the model is  fixed by the number of observations, and including more frequency channels increases also the complexity of the foreground emission to describe. We thus find a balance between having enough observations to allow good separation and reducing foregrounds complexity. \\
In the present configuration it is not possible to disentangle the clustered CIB and dust.
A more refined configuration, that includes a zone approach in SMICA and thus exploits the different spatial distributions of dust and clustered CIB, could in principle separate them.   The interest of this, apart from studying the properties of dust and cCIB \citep{cib:2017MNRAS.466..286M}, is that  it could improve the quality of the recovered CMB spectrum. The foreground contamination that is not accounted for by the model results in an increase of the final mismatch. However it is possible that a fraction of it projects on the CMB component. As we will see later, this can be checked with the aid of simulations.\\
The spectrum and emission law of dust and cCIB are freely fitted in each bin.  Instead, we impose some
constraints on the point sources part of the model, by making use of the physical knowledge we have.
Their spectra are constrained to be flat, consistently with the prediction that the point
sources can be modeled as shot noise. We expect that at the extrema of our frequency range only one population is clearly detected. This could induce the algorithm to find non-physical values for the emission law of the subdominant population. For this reason, we constrain the columns of the mixing matrix $\bold{A}$ relative to point sources to take only positive values, by fitting at each frequency the exponent of an exponential. Apart from positivity, we make no further assumption on the emission law shape. This configuration allows us to recover the joint emission law of point sources. It is not possible to disentangle the emission
of the two populations, since there is an intrinsic degeneracy between components
that have the same shape of the angular power spectrum \citep{Delabrouille:2002kz}. For this reason, throughout the text
we present results on the joint point sources emission.\\
The CMB angular power spectrum is freely fitted in each bin. Its emission law, with the calibration correction factors obtained by SMICA is instead fixed.\\
This refined model is
more useful for a physical understanding of the foregrounds, and answers the issue stated above of separating the CMB and point sources, which dominate at scales where the noise becomes important and are thus difficult to characterize. Without a dedicated model for
point sources, it is not possible to know what is their contribution  to the total
foreground level, hence it is not possible to correctly remove them from the CMB. In this sense point sources are degenerate with the CMB emission at small scales. We must note that the extra
information gained on point sources comes at the cost of increasing the mismatch between the data and the
proposed model  with respect to an unconstrained model.


\section{Data}\label{sec:data}
In this analysis, we use both simulations and Planck 2015 half-mission data.
\subsection{Simulations}\label{sec:simu}
\begin{figure}
\centering
\includegraphics[width=\hsize]{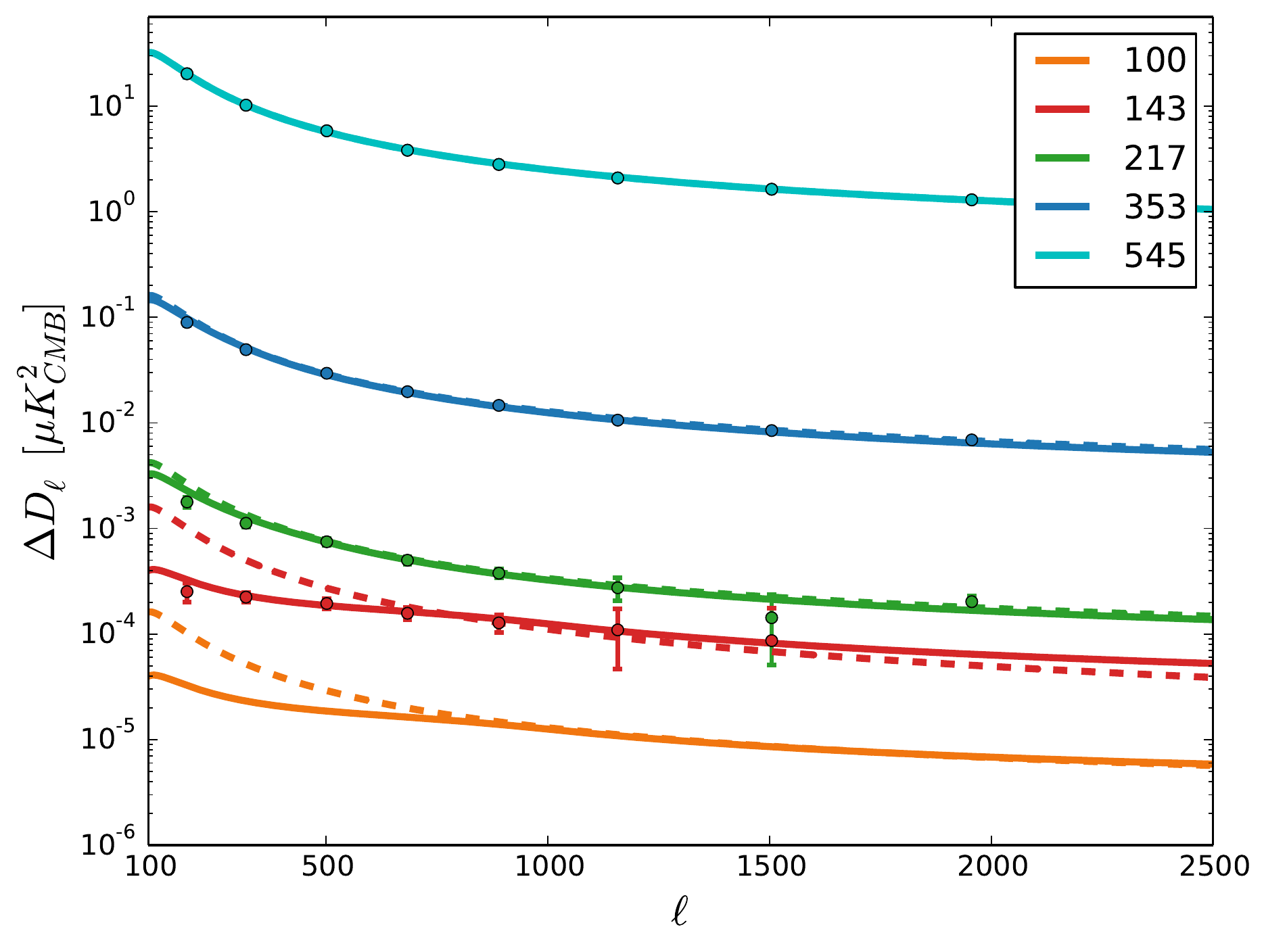}
\caption{ Angular power auto-spectra of the clustered CIB component used in simulations.
The dashed line represent the 1D clustered CIB spectra, obtained by fitting data points to a polynomial at 545 GHz and scaling this template to lower frequencies. The solid line represent the ND clustered CIB, obtained by separately fitting data points at each frequency. Overplotted data are taken from Table D2 in \citet{P13_cib}. No data points are available for the 100 GHz clustered spectrum, which is taken to be one order of magnitude less than the 143 GHz clustered spectrum. 
Spectra have been corrected for shot noise contribution in order to match data points.}\label{fig_cib}
\end{figure}
In order to test our model, we construct simulations of sky observations at the frequencies of interest,  which are a subset of the Planck HFI 
frequencies: 100, 143, 217, 353 and 545 GHz. For our main analysis we do not consider the 857 GHz channel, even though we also build simulations for this frequency: more  details about this choice are given in Sect. \ref{sec:857}.
The astrophysical emissions we consider are the CMB, the thermal dust and two extragalactic point sources populations, 
 the radio and the infrared ones.
For the latter, we simulate the clustered as well as the shot noise emission.\\
In order to better test our model with respect to extragalactic contamination, we produce three sets of simulations. They all  contain CMB, dust, radio point sources,  infrared point sources, clustered CIB and noise, but the properties of these signals differ in each set. We introduce here the general idea of the three different simulation sets. Technical details on how we build the different foreground components are given later.
The three simulations sets are:
\begin{itemize}
 \item SET1: these simulations have an idealised foreground content.  All the foregrounds are simulated as a single template rigidly scaled through frequency. We refer to foregrounds of these simulations as 1D or unidimensional, since their contribution in all auto-spectra of sky maps can be described by a single angular power spectrum rescaled in frequency, and they present no decoherence in the cross-spectra. Values for the angular power spectra of point sources and the clustered CIB at $\ell=3000$ are given in Table~\ref{tab:Simu}. The angular power spectra of the clustered CIB component are plotted in Fig.~\ref{fig_cib};
 \item SET2: these simulations include one foreground source with some frequency decoherence.  Galactic dust and the two point source populations are simulated as a 1D template each, which is rigidly scaled through frequency  (refer to Table~\ref{tab:Simu} for point sources). The clustered CIB presents some frequency decoherence, i.e.:
\begin{equation}\label{cib_decoherence}
C_{\ell}^{CIB ~  \nu_1 \times \nu_2} = \alpha_{\nu_1 \times \nu_2} \sqrt{C_{\ell}^{CIB ~   \nu_1} C_{\ell}^{CIB ~  \nu_2}},
\end{equation} 
where the coefficients $\alpha_{\nu_1 \times \nu_2} \leq$1  are taken from \citet{P13_cib} and are  reported in Table~\ref{tab:Decorr}. The angular power spectrum shape is modeled on observational estimates: 
 the shape of the angular power spectra at each frequency differ slightly. The power spectra at different frequencies are presented in Fig.~\ref{fig_cib}. We refer to this CIB component as ND or multidimensional;
 \item SET3: these simulations have the most realistic foreground content. The two point source populations are simulated as a 1D template each, which is rigidly scaled through frequency  (again refer to Table~\ref{tab:Simu}). The clustered CIB is simulated as in SET2. The dust component presents spectral index and dust temperature variability on the sky, using results from \citet{P15_gnilc}. We refer to this dust component as ND or multidimensional.
\end{itemize}
These three sets are labeled SET1, SET2 and SET3 throughout this work.
The SET2 and SET3 cases are studied since observations show that there could be a partial decoherence through frequency of the CIB emission \citep[Sect.6.2]{P13_cib}, 
this effect being mostly evident at the two lowest frequencies 100 and 143 GHz. The third case also includes a realistic dust representation, which takes into account the inhomogeneous dust properties on the sky.
Both are important tests since the SMICA method assumes no frequency decoherence or variability of the spectral index for the unidimensional sources: in the SMICA model this variability is accounted for as an increase of the dimensionality of the source. However the model has a maximum number of dimensions fixed by the number of observations.\\
In order to reproduce the Planck half-mission maps used in this analysis, for each simulation  we produce $2$ maps for each frequency, 
and each couple of maps at same frequency has identical astrophysical content but a different realization of white Gaussian noise.
We produce N=30 simulations for each set.\\
\par\smallskip
\noindent \textbf{Building the components}
The CMB component is simulated from a theoretical CMB temperature angular power spectrum using the HEALPix tool \citep{Gorski:2004by}. The power spectrum is obtained using the code CosmoMC, with the following set of input cosmological 
parameters: $H$ = 67.31, $\tau$ = 0.078, $\omega_b$ = 0.02222, $\omega_c$ = 0.1197, $n_s$ = 0.9655, $\ln( 10^{10} A_s)$ = 3.089, $y_{He}$ = 0.24 and $m_{\nu}$ = 0.06 eV.\\
There are two different thermal dust components: one is a single template scaled through frequency (SET1 and SET2 simulations), while the other presents more complex features (SET3).
The former, labeled ``1D'', is the thermal dust map at 545 GHz delivered by the {\it Planck Collaboration} 
\citep{P15_compsep_for}, which we choose in order to have a realistic spatial distribution. This template is scaled through frequency according to the grey-body law
described by Eq. (\ref{dust_em}) with T=19.4 and $\beta$=1.6 \citep{P15_gnilc}. 
 Note that this template is partially contaminated by residual CIB emission \citep[Sect. 4]{P15_compsep_for}, which makes the dust contribution of this template at high $\ell$ higher than the real dust contribution. This makes the fit of the high-$\ell$ components slightly more difficult for SMICA. 
Due to the fact that thermal dust and the clustered CIB have similar emission laws, the presence of a residual contamination of CIB in the small scales of the thermal dust template map is not to be excluded , i.e., the small scale power of this template could be higher than the real dust distribution.
The latter, labeled ``ND'',  is simulated using the GNILC model maps for the spectral index $\beta_d$, the dust temperature, and the opacity, obtained as described in \citet{P15_gnilc}. They are combined through Eq. (\ref{dust_em}) to produce a dust map at each $\nu$.\\
For the extragalactic content, that is point sources and clustered CIB, we base ourselves on \citet{P13_cib}, which provides estimates for the radio and infrared point sources shot noise levels,  the angular power spectra of CIB emission and its decoherence coefficients  at Planck frequencies. 
Shot-noise levels are given at all the frequencies of interest of this paper, and we therefore use them for point source simulations. We model the two point source populations as two realisations of shot noise maps, i.e., with a flat angular spectrum.
 The amplitudes of the shot noise power are taken from Table 6 and 7 in \citet{P13_cib} and are summarized in Table \ref{tab:Simu}.\\
CIB spectra and decoherence coefficients are given by the Planck analysis for all frequencies except 100 GHz: 
we choose for this channel values one order of magnitude lower than the 143 GHz estimates. The CIB angular power spectra reported in Table D2 of \citet{P13_cib} contain both the clustered and shot-noise contribution: the latter is subtracted to obtain clustered CIB templates.  
\begin{table*}
\caption{\label{tab:Simu} 
Simulation parameters for point sources and clustered CIB  as $C_{\ell=3000}$ levels in Jy$^2$/sr. }
\centering
\begin{tabular}{|c|ccc|}
\hline
\hline
      	& Radio Point Sources 	&  	IR Point Sources	& 1D clustered CIB 	\\
\hline
100 	& 8.48					&   0.150				& 0.136 				\\
143 	& 6.05					&   1.20				& 3.43				\\
217 	& 3.12					&   16.0 				& 14.4					\\
353 	& 3.28					&   225 		  		& 209				\\
545 	& 2.86					&   1454 				& 1550 					\\
857 	& 4.28				   	&   5628 				& 5397 				\\
\hline
\end{tabular}

\end{table*}

\noindent To produce the clustered CIB component maps at each frequency we compute the covariance matrix $R_{\ell}^{CIB}$ of CIB auto- and cross-angular power spectra. More specifically:
\begin{itemize}
\item for SET1, i.e., the 1D clustered CIB, we fit a polynomial to the data points of the 545 GHz power spectrum, and all the other auto- and cross-spectra are obtained by scaling this template. Scaling coefficients for auto-spectra are obtained from \citet{P13_cib}, while for cross-spectra we use Eq. (\ref{cib_decoherence}) with $\alpha_{\nu_1 \times \nu_2}$ = 1. $C_{\ell=3000}$ values are reported in Table \ref{tab:Simu};
\item for SET2 and SET3, i.e., the ND clustered CIB, at each frequency we  fit a polynomial to the data points of the auto-spectra given in \citet{P13_cib}, and we extrapolate to higher $\ell$ when necessary. Only the auto-spectra are used, while cross-spectra are derived via Eq. (\ref{cib_decoherence}).  The decoherence coefficients of angular power spectra between different frequencies are detailed in Table \ref{tab:Decorr}.
\end{itemize}
The auto-spectra for both cases are presented in Fig.~\ref{fig_cib}.
Once the covariance matrix $R_{\ell}^{CIB}$ is constructed, the procedure for obtaining spherical harmonics is the same for both cases.
We build the vector $\bold{x}_{\ell m}$, whose entries $x_{\ell m}^{i}$ are sets of spherical harmonics coefficients 
drawn from the normal distribution,
\begin{equation}
 x_{\ell m}^{i} \sim \mathcal{N}(0,1),
\end{equation}
where  $i=1,2,...N$, and $N$ is the number of frequencies we use.
We then obtain spherical harmonics for the CIB as: 
\begin{equation}
 x_{\ell m}^{CIB}=  Z_{\ell}^{CIB} \bold{x}_{\ell m}
\end{equation}
where $Z_{\ell}^{CIB}$ is the square root of the clustered CIB covariance matrix $R_{\ell}^{CIB}=Z_{\ell}^{CIB}Z_{\ell}^{CIB}$. 

\begin{table}
\caption{\label{tab:Decorr} 
Decoherence coefficients for the ND clustered CIB.  }
\centering

\begin{tabular}{|c|cccccc|}
\hline
\hline
      	& 100 	& 143 		& 217 	 	& 353	 	& 545	 	& 857	\\
\hline
100 & 1		&	-	&	-	&	-	&	-	& -	\\
143 & 0.99 	&	1	&	-	&   -     	&   - 		& -	\\
217 & 0.78    	&	0.78	&	1	&	-	&   - 		& -	\\
353 & 0.54      &  	0.54	& 	0.91    &	1	&	- 	& -		\\
545 & 0.51    	&  	0.51 	& 	0.90    & 	0.983	&	1	& -	\\
857 & 0.45    	&  	0.45 	& 	0.85    & 	0.911	&	0.949	& 1	\\

\hline
\end{tabular}

\end{table}

\noindent In order to build simulations, the CMB and foregrounds maps are added with their respective amplitude for each frequency and then smoothed with their respective beam window function\footnote{\label{beam}provided in Planck's RIMO, which can be downloaded from the {\it Planck Legacy Archive} http://pla.esac.esa.int/pla/}.
By construction, there is no correlation between the foregrounds and the CMB.\\
The instrumental noise is simulated at the map level as white Gaussian noise.
Noise amplitudes are determined using Planck noise simulations as provided at NERSC\footnote{http://crd.lbl.gov/cmb-data}.

\subsection{Planck data}\label{sec:data2}
We use data maps from the 2015 full Planck release and we select the two half-mission maps at each frequency between 100 and 545 GHz. Half-mission maps are data split
obtained by dividing the full mission time-ordered data into two halves.
The maps are degraded to a lower resolution of Nside = 1024 using {\it HEALPix}. 

\subsection{Masks and binning}\label{sec:masks}
\begin{figure}
\centering
\includegraphics[width=\hsize]{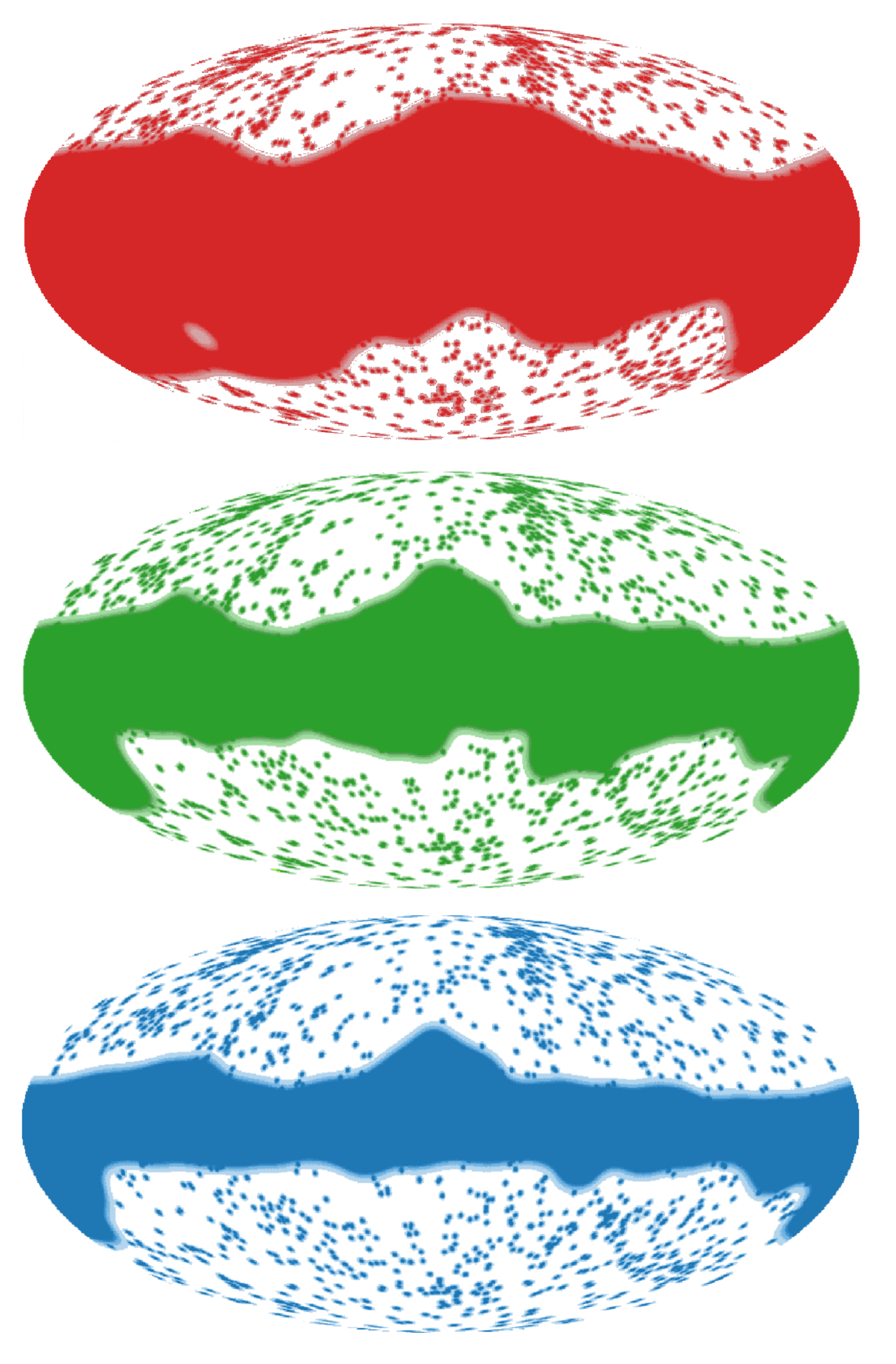}
\caption{Apodised masks used in this analysis. The retained sky fractions $f_{sky}$=0.3, 0.5, 0.6 are shown in red, green and blue respectively. The shaded region is the apodised part. }\label{FigMask}
\end{figure}
\begin{figure}
\centering
\includegraphics[width=\hsize]{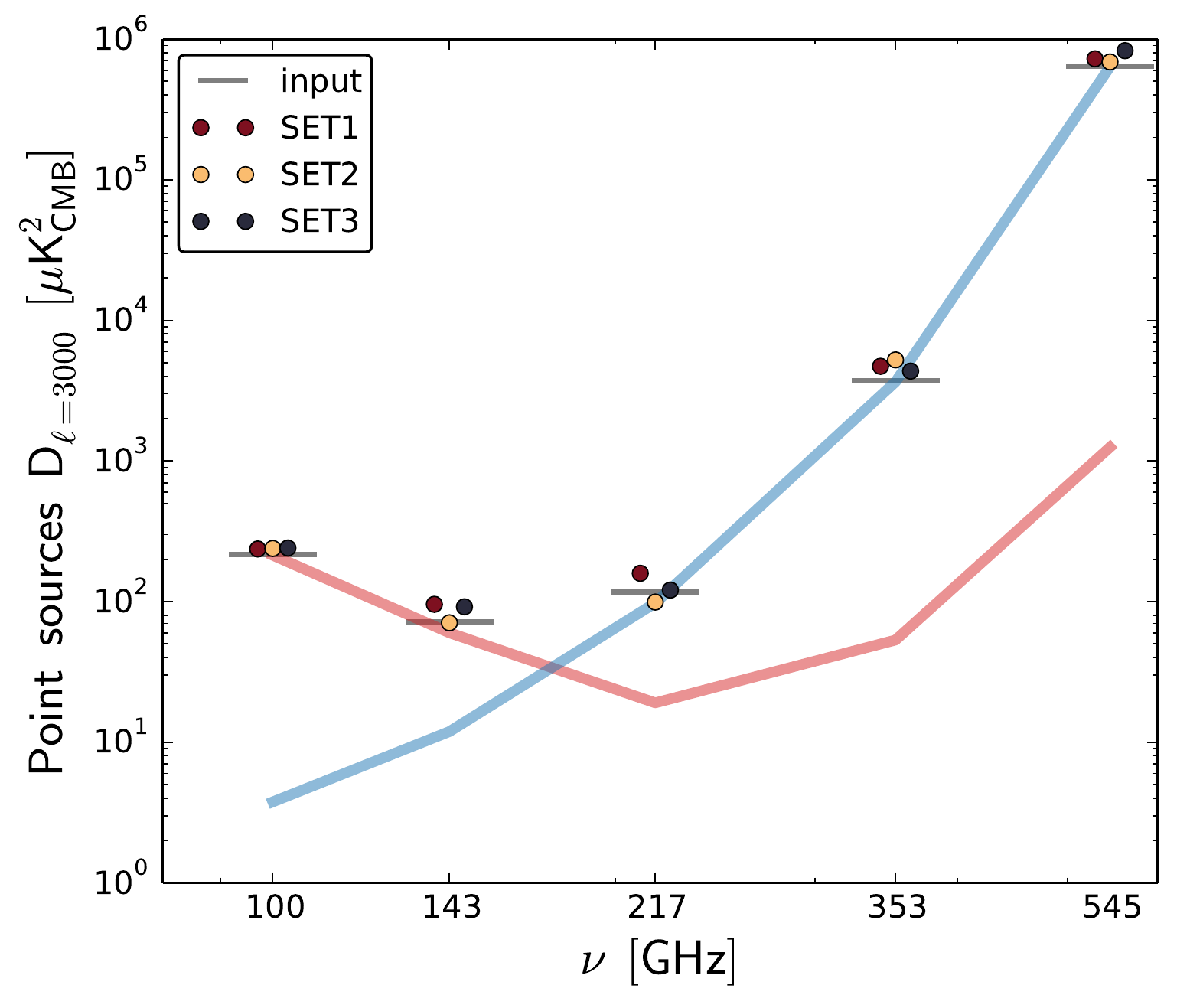}
\includegraphics[width=\hsize]{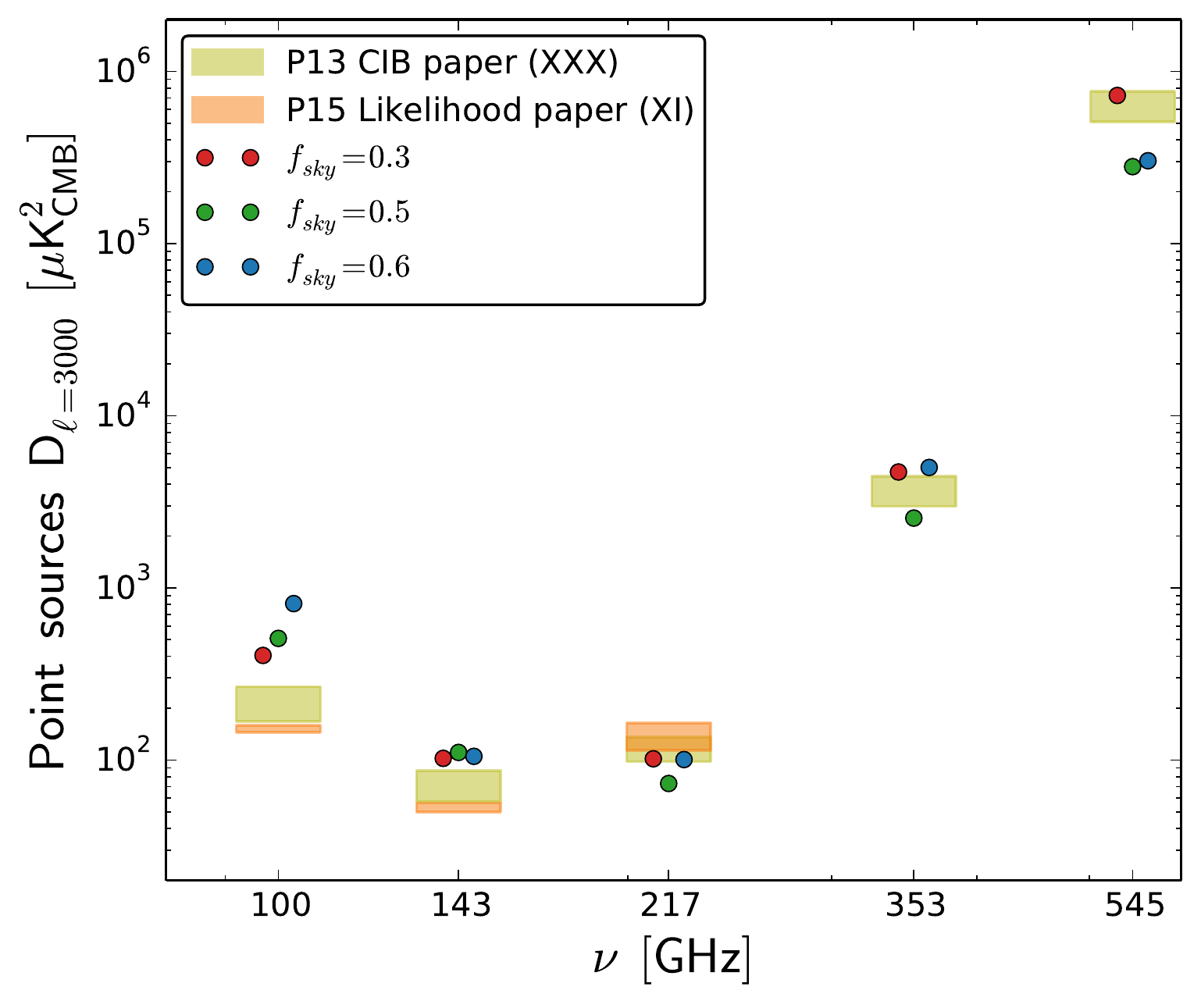}
\caption{Combined (infrared and radio) shot noise point sources $D_{\ell}$ power at $\ell=3000$ obtained from the fit. {\it Top panel}: simulations average of point sources at $f_{sky}=0.5$, 
shown in dark red for SET1 simulations, yellow for SET2 and black for SET3. The red and blue band show the simulations 
input for the radio and infrared point sources respectively, while the grey horizontal line at each $\nu$ represents the joint point source input. 
{\it Bottom panel}: point sources for the three different masks of $f_{sky} = 0.6, 0.5, 0.3$ in blue, green and red respectively. 
The yellow and orange bands represent the expected shot noise point source contribution estimated in \citet{P13_cib} and 
\citet{P15_lkl}, where the width of the coloured band represents the error on the expected value. \citet{P15_lkl} gives expected values 
for the three low $\nu$ channels only.}\label{fig_ps}
\end{figure}
\begin{figure}
\centering
\includegraphics[width=\hsize]{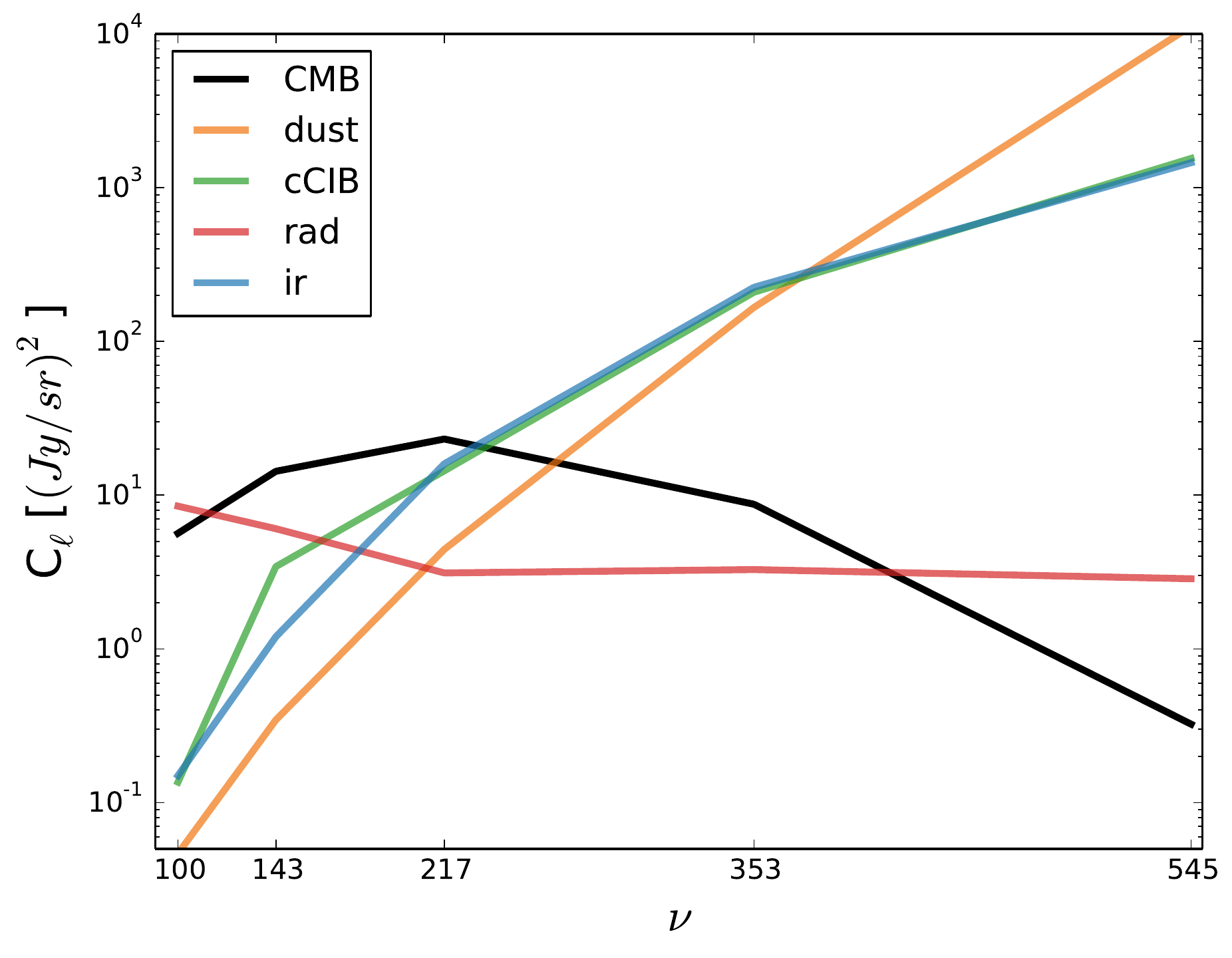}
\caption{Input spectral behaviour of dust, clustered CIB, infrared and radio point sources of SET1 simulations at $f_{sky} = $ 0.5, plotted as $C_{\ell=3000}$.  
Radio and infrared point sources are labeled ``rad'' and ``ir'' respectively. 
Infrared point sources, cCIB and dust, plotted in blue, green and orange respectively, 
present similar emission laws.
CMB black-body emission law, which is not fitted for, is plotted in black.} \label{fig_mixmat}
\end{figure}
In order to reduce the foreground contamination, the central regions of the sky are masked.
Masks are produced as a sum of a galactic and a point source part. We use a set of three masks with the same point source masking but different galactic coverage.
The masks used are shown in Fig.~\ref{FigMask} and have effective $f_{sky}$ = 0.3, 0.5, 0.6. More details on the masks preparation are given in Appendix \ref{appa}.\\
Since SMICA works with spectral covariance matrices, angular power spectra between all couples of maps are calculated with the {\it PolSpice} \citep{Chon:2003gx} package. 
Using the {\it PolSpice} routine, we correct the resulting power spectra for the point spread function of the instrument using the beam window functions provided by the full Planck 
release\footnotemark[3], for the pixel window function and for the mask leakage.
All the angular power spectra are binned uniformly with $\Delta \ell = 15$. With these spectra, and following the procedure detailed in Sect. \ref{sec:data-splits}, we build at each bin a $5\times 5$ covariance matrix $\hat{R}_{\ell}^{split}$.
We work on the range $\ell=[100,2500]$: we neglect in this analysis the large angular 
scales $\ell < 100$, where dust has complex features that can not be described by a bidimensionnal component. Also, we  limit our 
analysis around $\ell \sim 2500$, since for higher multipoles noise becomes dominant.\\
The Planck maps are slightly decalibrated among each other. Similarly to what is done in the Planck analysis \citep{P13_compsep}, 
we perform a dedicated free fit in the multipole range of the first and second peak to recover calibration factors. 
We use relative calibration correction factors with respect to 143 GHz: $y_{cal}$=[1.00079, 1., 1.0029, 1.008, 1.0174] 
for the five channels between 100 and 545 GHz.

\section{Testing the method}\label{sec:test}
 We detail here the analysis and fitting procedure to obtain the CMB power spectrum.
We test this method on simulations first and Planck 2015 temperature data then.
The spectra recovered from simulations and data are used to estimate cosmological parameters, 
which are presented in Sect. \ref{sec:cosmo}.

\subsection{Simulations analysis}\label{sec:analis_simu}
\begin{figure*}
\centering
\begin{minipage}{.45\linewidth}
\includegraphics[width=\textwidth]{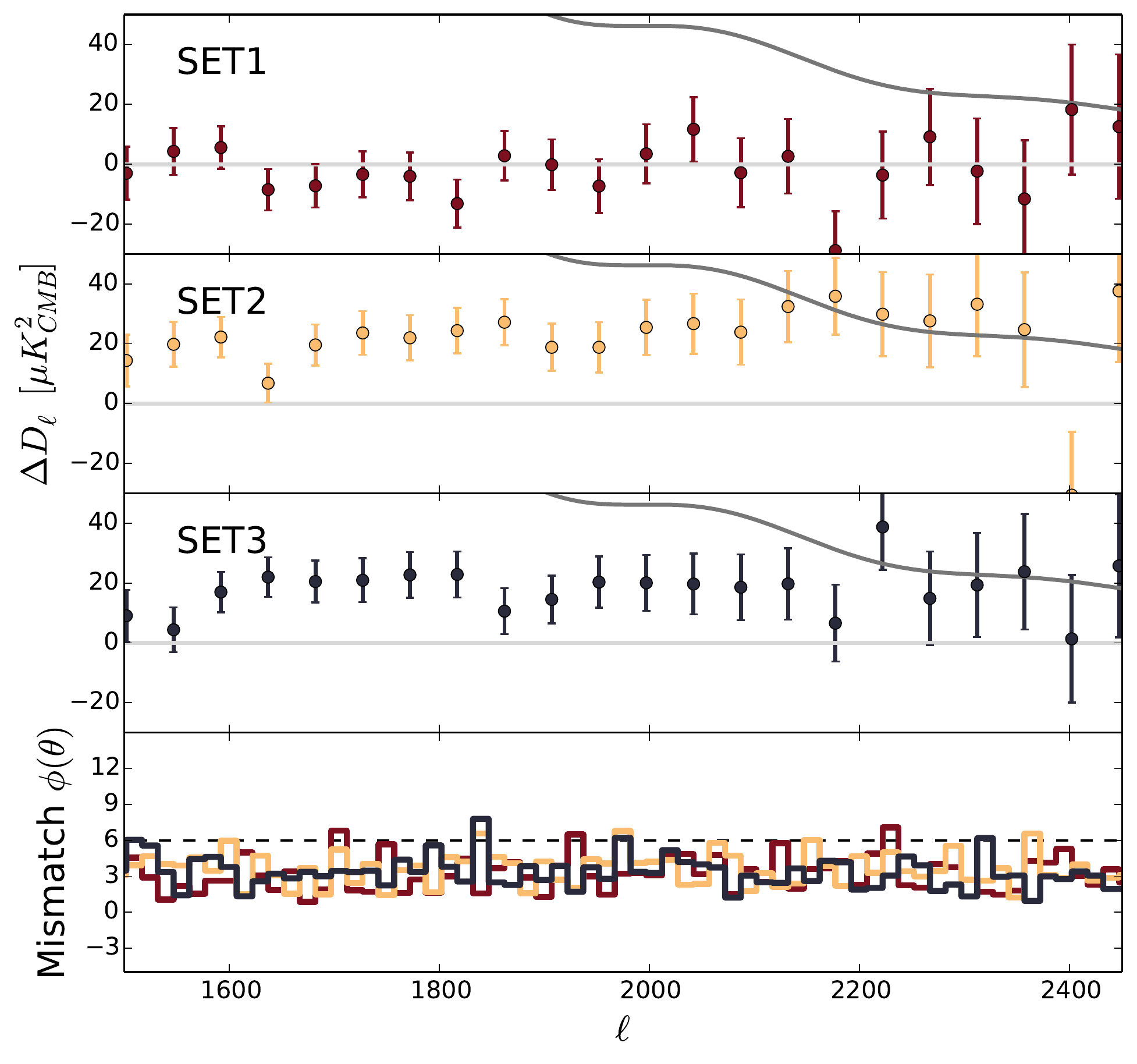}
\end{minipage}
\hspace{.05\linewidth}
\begin{minipage}{.45\linewidth}
\includegraphics[width=\textwidth]{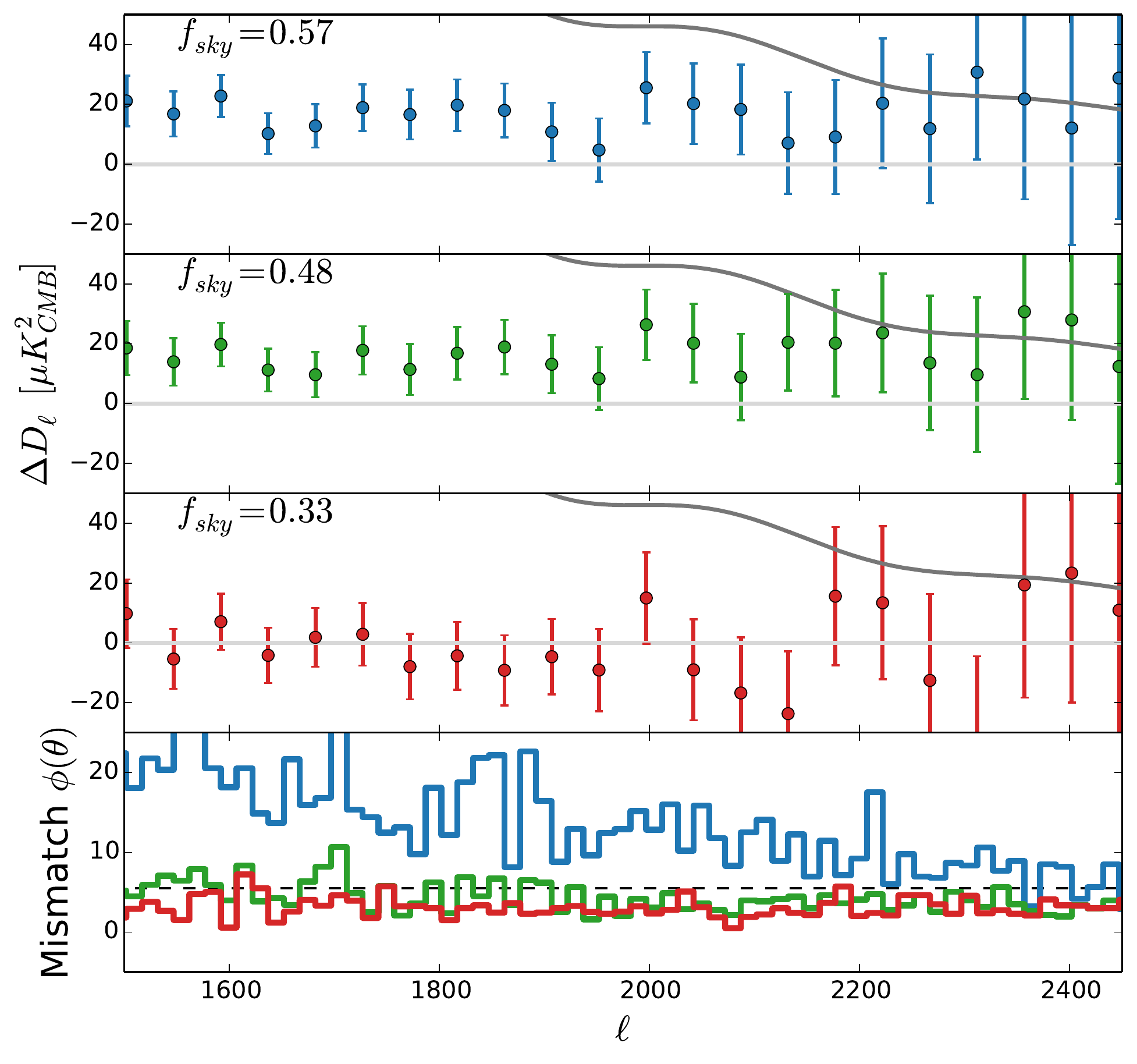}
\end{minipage}
\caption{Top three panels show the residuals in $D_{\ell}$ between the fit results and the theoretical CMB power spectrum. Dark grey line shows the 
theoretical CMB spectrum downscaled by a factor 0.2 for readability.  Bottom panel shows the mismatch between the model and the data after the fit as defined by 
Eq. (\ref{mismatch}) and the thin dashed line shows the expected mismatch per bin. Only one point every three bins is displayed.
{\it Left panel:} Filled dots show differences between CMB spectra obtained from fit on simulations with respect to input CMB spectrum at each bin, 
at $f_{sky}=0.5$, shown in dark red for SET1, yellow for SET2 and black for SET3. 
{\it Right panel}: Filled dots show differences between CMB spectra obtained from the fit on Planck 2015 half-mission data with respect to the 
Planck best fit $\Lambda$CDM Plik spectrum.}\label{fig_cmb}
\end{figure*}
While the simulated foregrounds can not reproduce the full complexity of real data foregrounds, a study on simulations is a good test for understanding to which degree
we can recover the point source signal and the CMB angular power spectrum.
We process the three simulation sets with the foreground model described by Eq. (\ref{SMICA:model}). For SET1, since foregrounds are all 1D, we constrain the $P_{\ell}^{dust+cCIB}$ component to be diagonal.\\
In the top panel of Fig. \ref{fig_ps} we show the recovered shot noise point source signal for the average of all simulations of each set at $f_{sky}$ = 0.5. \\
We show results for the intermediate $f_{sky}$ value, but we observe no mask dependence in 
the recovered point source emission.
We observe that the model is capable of recovering closely, up to small offsets, the point sources input for all the 3 cases.
The SET1 case, which has all 1D foregrounds and is therefore an ``ideal'' test case for SMICA, presents a small offset in the three central frequencies. 
This is not surprising: even though the foreground content corresponds exactly to the SMICA model, the clustered CIB, the infrared shot-noise and the galactic dust have similar emission laws, 
and the corresponding columns of the matrix $A$ are almost proportional, as seen in Fig.~\ref{fig_mixmat}. 
This is far from ideal for ICA methods,  since it limits the identifiability of the sources. Due to this we expect an exchange in power between dust, cCIB and infrared point sources.
The offsets in SET2 and SET3 simulations are instead likely due to the fact that the model is incapable of representing the foregrounds complexity due to its limited dimensionality.\\
In the left panel of Fig. \ref{fig_cmb} we show residuals of the high-$\ell$ tail of the fitted CMB angular power spectrum with respect to the theoretical input. As seen from this Figure, the residual is at most one fifth of the CMB power at $\ell \geq 2200$.
A residual contamination is present on average for the SET2 and SET3 cases only. 
The misevaluation of the clustered CIB contamination can be one  source of bias in the CMB power spectrum estimation.
The SMICA method assumes full correlation of all components through frequency. A partial decoherence of a component, as for example in SET2 
for the clustered CIB, means that its spectral behaviour must be described by a multidimensional component. 
For galactic dust and clustered CIB, we have a 2D component describing both of them at the same time.  
While angular power spectra are fitted in each bin, the mixing matrix $A$ is global: galactic dust and clustered CIB emissions, which are important at low and high multipoles respectively, compete for the columns  of this matrix.
As a consequence, complex features in these two emissions can not be fully accounted for. We expect that a part of the CIB and dust contamination projects onto the
CMB, resulting in an offset with respect to the input spectrum, as shown in the left panel of Fig. \ref{fig_cmb}. \\
We can see that such a contamination is not detectable as a considerable increase in the mismatch, while it is clearly visible in the 
CMB residuals. 
Results in Fig. \ref{fig_cmb} are presented for $f_{sky}$ = 0.5, 
but no significant trend with sky fraction is visible in most simulations.
We see that the observed mismatch is lower than the expected value. This happens because of the peculiar statistical properties 
of the empirical covariance matrices used in this work, as described in Sect.~\ref{sec:data-splits}. 
The value of the mismatch does not correspond to the number of degrees of freedom $\nu$, 
which we plot anyway as a visual reference of the order of magnitude of the expected mismatch (see Sect.~\ref{SMICA:specmatchcrit} for more details).

\subsection{Data analysis}\label{sec:data_analis}
\begin{figure}
\centering
\includegraphics[width=\hsize]{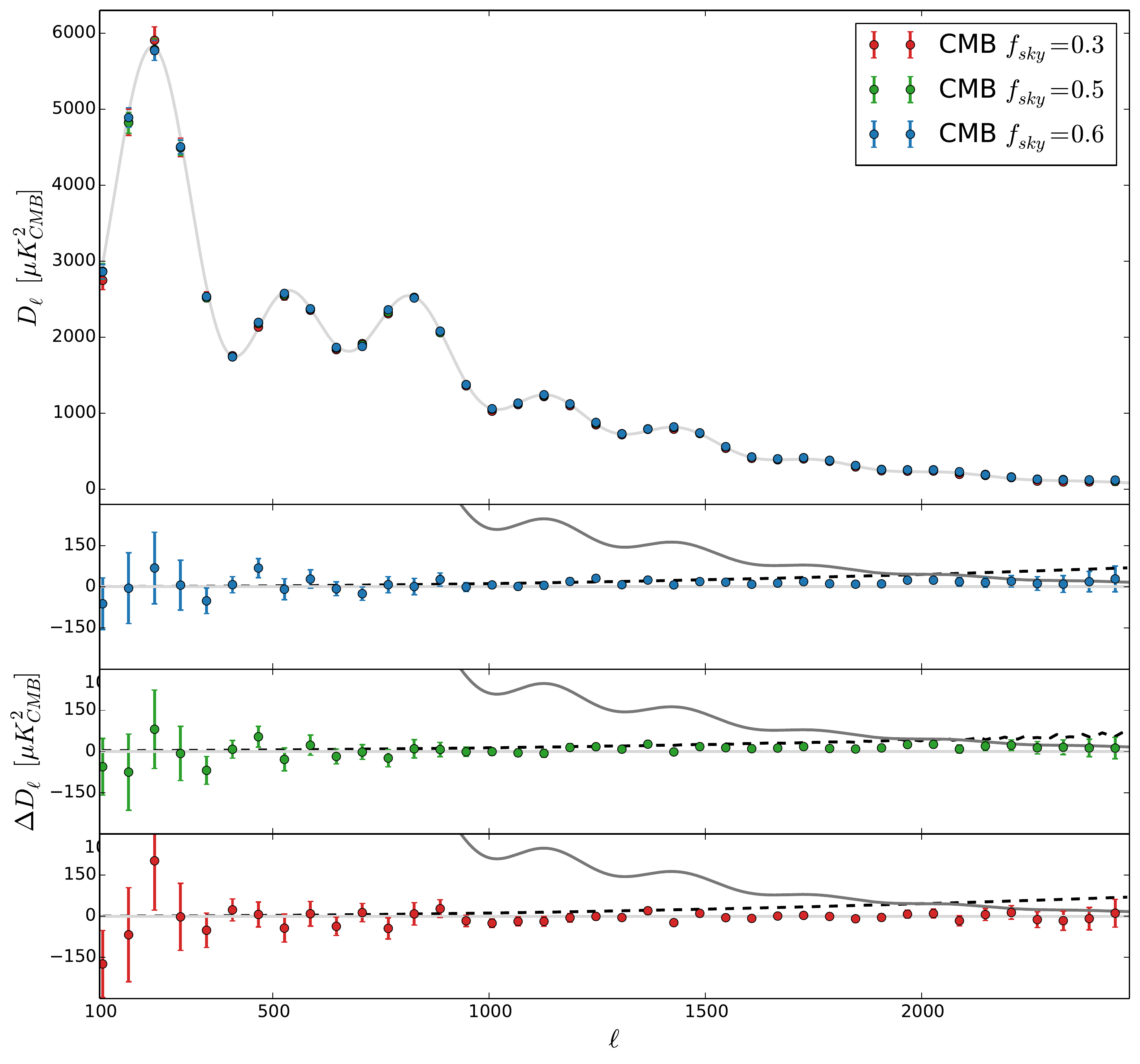}
\caption{CMB angular power spectrum obtained from the SMICA fit of the model to the three different data sets used, corresponding to the masks 
with $f_{sky}$= 0.3, 0.5, 0.6 (top panel). In grey we show the best fit theoretical $\Lambda$CDM spectrum obtained with Plik. Three bottom panels show the
residuals between theory and data for the three cases, together with the theoretical value of the angular power spectrum scaled to 20\% power in grey. The black dashed line shows the total contribution of extragalactic foregrounds at 217 GHz. To enhance readability, only one point in four is plotted. }\label{fig_data_cmb1}
\end{figure}

We fit a model as described in Section \ref{sec:smica3}.
The obtained CMB angular power spectrum is presented in Fig.~\ref{fig_data_cmb1} for the three different masks, 
while right panel of Fig. \ref{fig_cmb} shows residuals with respect to the 
reference CMB Planck spectrum at high-$\ell$. 
The reference Planck spectrum is the theoretical $\Lambda$CDM spectrum obtained from best fit parameters of the Planck 2015 Plik likelihood exploration.
Error bars are derived with the Fisher matrix.\\
We observe in Fig. \ref{fig_cmb} that the results obtained for the CMB are in good agreement between the three different masks. 
We can see an increasing level of residual contamination, for increasing sky fraction. While 
this trend is not seen in simulations, we expect such a behaviour in real data since the foreground complexity increases. 
As observed in simulations, we expect that the model cannot fully capture dust and cCIB emission. 
Also, our simulations contain two point source populations perfectly correlated through frequency. 
While this is a good approximation, it might not represent the full extent of contamination produced by background galaxies. 
Another problem is the similar emission law between dust and CIB, which cannot be fully captured by the model; due to this, 
a fraction of the foreground contamination projects on the CMB and on the mismatch
between the model and the data.
We see that the mismatch is much higher than the mismatch found in simulations, in particular for the smallest mask and at low multipoles, where the thermal dust behaviour becomes more complex.\\
In the bottom panel of Fig. \ref{fig_ps} we show the recovered point sources amplitudes for the three masks at $\ell=3000$. 
Results for $f_{sky}$ = 0.3, 0.5 are in good agreement with each other and with the expected amplitude as estimated by the 
{\it Planck Collaboration}. The $f_{sky}$ = 0.6 results show an offset at the highest and lowest frequencies: again the model fails
to fully represent the foregrounds complexity. We expect point sources estimates at smaller $f_{sky}$ to be more accurate, since the galactic contamination is lower. The offset of point source emission law is related to the offset in the CMB power spectrum, but cannot fully explain it. Forcing point sources emission law to the result obtained for the largest mask, i.e., to a value closer to the  expected one, reduces only slightly the mismatch and the CMB bias.

\subsection{Using 857 GHz}\label{sec:857}
The number of channels used is directly related to the dimensionality of the foreground model. Including more observations allows
for a higher dimension, but also adds new features in the data which need to be described. We choose to exclude low frequency observations from our 
analysis since this would include synchrotron and free-free emission and thus increase the Galactic foreground complexity. 
We also choose to exclude WMAP 94 GHz observations since they have a lower resolution than Planck data and this would oblige 
us to use a smaller $\ell$ range. \\
Higher frequency observations could in principle be useful since they contain mainly dust, infrared point sources and clustered CIB. However frequency decoherence 
of foregrounds makes the effective impact of high frequency channels negligible.
We present in this section results on SET3 simulations and Planck data when adding the 857 GHz channel.  For the analysis on data, the masks are adapted by adding
point sources detected in the 857 GHz maps, but effective sky fractions are substantially unchanged. 
The fitting procedure is the same as described in Sect. \ref{sec:analis_simu} and Sect. \ref{sec:data_analis}, with
the only difference that the $P_{\ell}^{dust+cCIB}$ part of the model in Eq. (\ref{SMICA:model}) has now three dimensions instead of two. \\
For simulations we see no evident difference in the SMICA fit between adding or not the 857 GHz channels. 
For data, the CMB power spectrum for $f_{sky}$ = 0.5, 0.6 is shown in Fig. \ref{fig_restat}. No improvement is seen with respect to the fit without the 857 GHz channel. 
While simulations show a good agreement between masks, the recovered point sources emission laws show an evident  bias at low frequencies $\nu \leq$ 217 GHz. This hints that a degree of decoherence is present between 857 GHz and lower frequencies shot noise emission. The mixing matrix columns reserved to point sources can not accommodate for both high and low frequencies, sacrificing the latter. 

\subsection{Without data-splits}\label{sec:nosplit}
\begin{figure}
\centering
\includegraphics[width=\hsize]{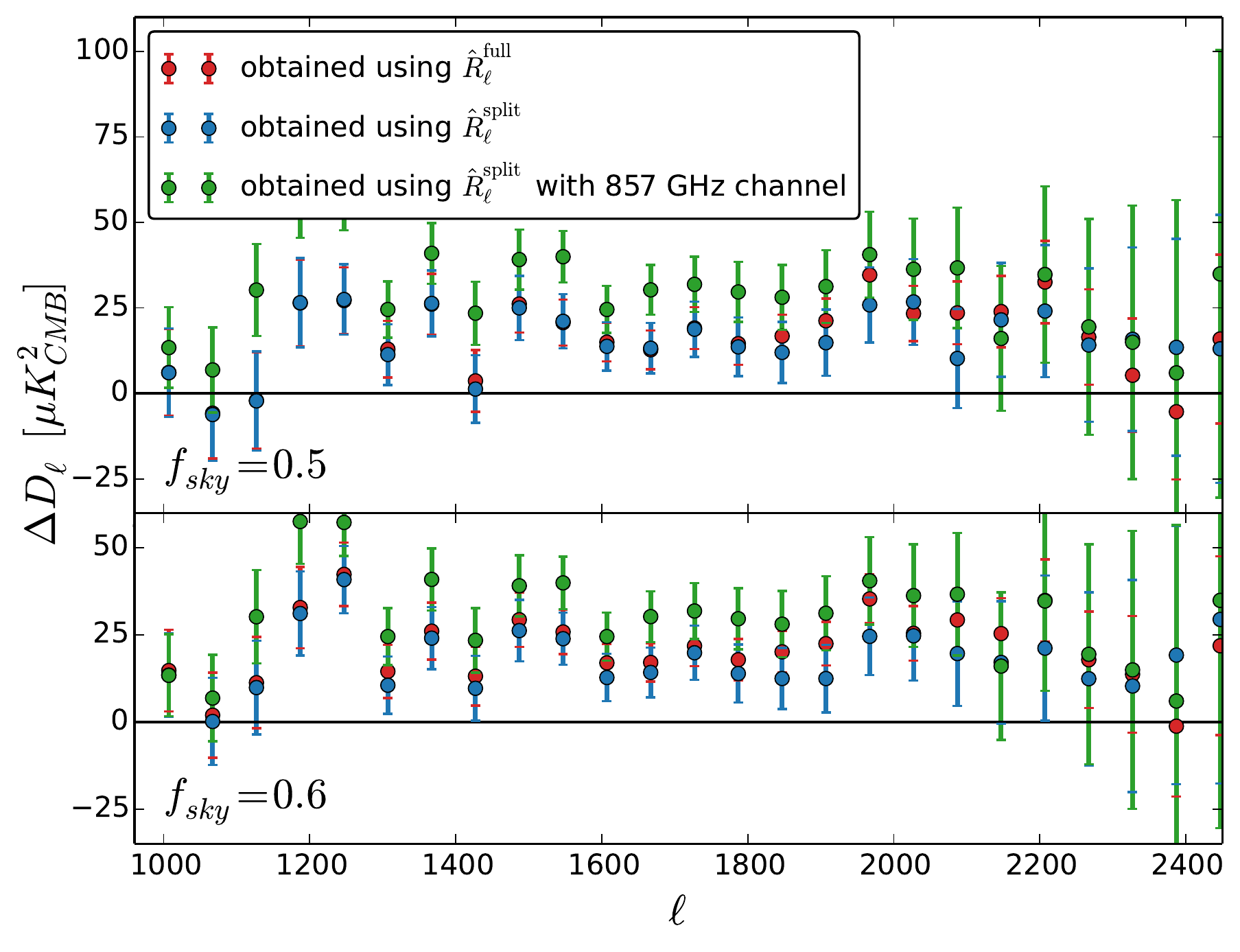}
\caption{CMB angular power spectrum $D_{\ell}$ residuals with respect to Planck theoretical best fit spectrum for three different SMICA configurations.
In blue we show the leading configuration of this paper using cross-spectra of data  splits, in red the one using 
cross- and auto-spectra as described in Sect. \ref{sec:nosplit} and in green the results obtained including 857 GHz channel observations, as detailed in 
Sect. \ref{sec:857}. Only one point every three bins is displayed. We show results for $f_{sky}$ = 0.5 in the top panel and $f_{sky}$ = 0.6 in the bottom panel. We note that the case which uses $\hat{R}_{\ell}^\mathrm{split}$ without the 857 GHz channel has an overall smaller residual with respect to the other two cases.}\label{fig_restat}
\end{figure}

The configuration described in Sect. \ref{sec:data-splits} tests covariance matrices built using data split cross-spectra only. 
A simpler configuration would be to use the full 
$2N \times 2N$ covariance matrix of auto- and cross-spectra, where $N$ is the number of frequency channels. This matrix is defined as:
\begin{equation}
\hat{R}_{\ell}^\mathrm{full}= \frac{1}{2\ell +1} \sum_m \bold{y}_{\ell,m}^\mathrm{full}  {\bold{y}_{\ell,m}^{full}}^T,
\end{equation}
where $\bold{y}_{\ell,m}^\mathrm{full}= [\bold{y}_{\ell,m}^{a},\bold{y}_{\ell,m}^{b}]$. The model used in this case is:
\begin{equation}\label{eq_model_full}
  R_\ell ( \theta )
  = 
  \begin{bmatrix}
    \bold{a} & F \\
  \end{bmatrix}
  \
  \begin{bmatrix}
    C_\ell^\mathrm{cmb} & 0 \\
    0 & P_\ell 
  \end{bmatrix}
  \
  \begin{bmatrix}
    \bold{a} & F \\
  \end{bmatrix}^T 
  + N_\ell
\end{equation}
where $N_\ell$ is the diagonal matrix containing the noise power spectra.
In this configuration the noise power spectra are part of
the fitted parameters.  This higher number of parameters to fit is compensated by the increased dimension of the
data matrix $\hat{R}_\ell$.\\
On Planck data, we show in Fig. \ref{fig_restat} that residuals for the cross-spectra only covariances $\hat{R}_\ell$ are lower than 
those obtained using the auto- and cross-spectra covariances $\hat{R}_\ell^\mathrm{full}$. 
We attribute this difference to the higher number of parameters to fit in the full matrix case. Also, an error in the noise estimation reflects on the
astrophysical part of the fit, and potentially on the CMB. Instead, in the configuration chosen for this study, noise spectra are known by construction and are not fit for, and thus they can
not bias the fit. The drawback in this case is that the estimated error bars depend on the noise ansatz (see Sect. \ref{SMICA:specmatchcrit}).
\begin{figure}
\centering
\includegraphics[width=\hsize]{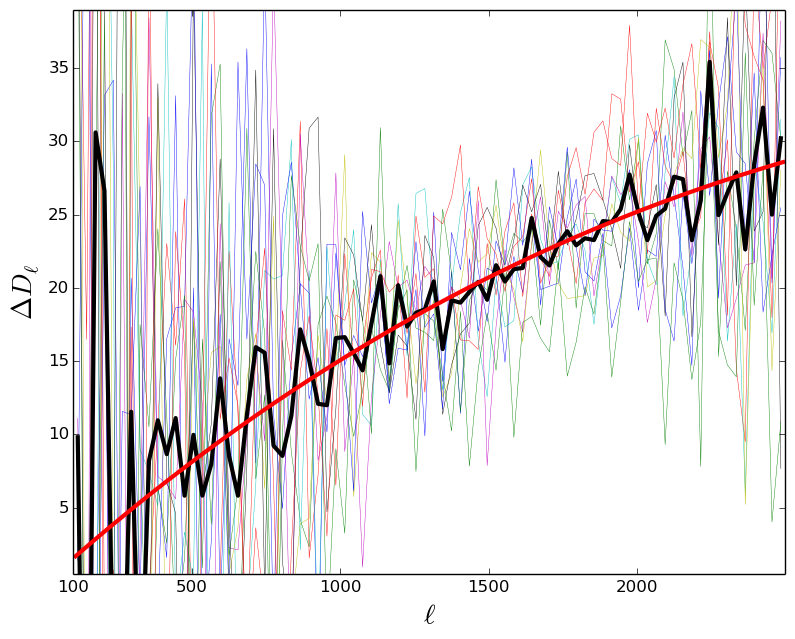}
\caption{ Difference between SMICA best fit and input maps CMB angular power spectra for 10 SET3 simulations. The average of the differences is plotted in black, while the chosen template for the likelihood is plotted in red.}\label{fig_template}
\end{figure}

\section{Cosmological parameters}\label{sec:cosmo}
\begin{figure*}
\centering
\includegraphics[width=\hsize]{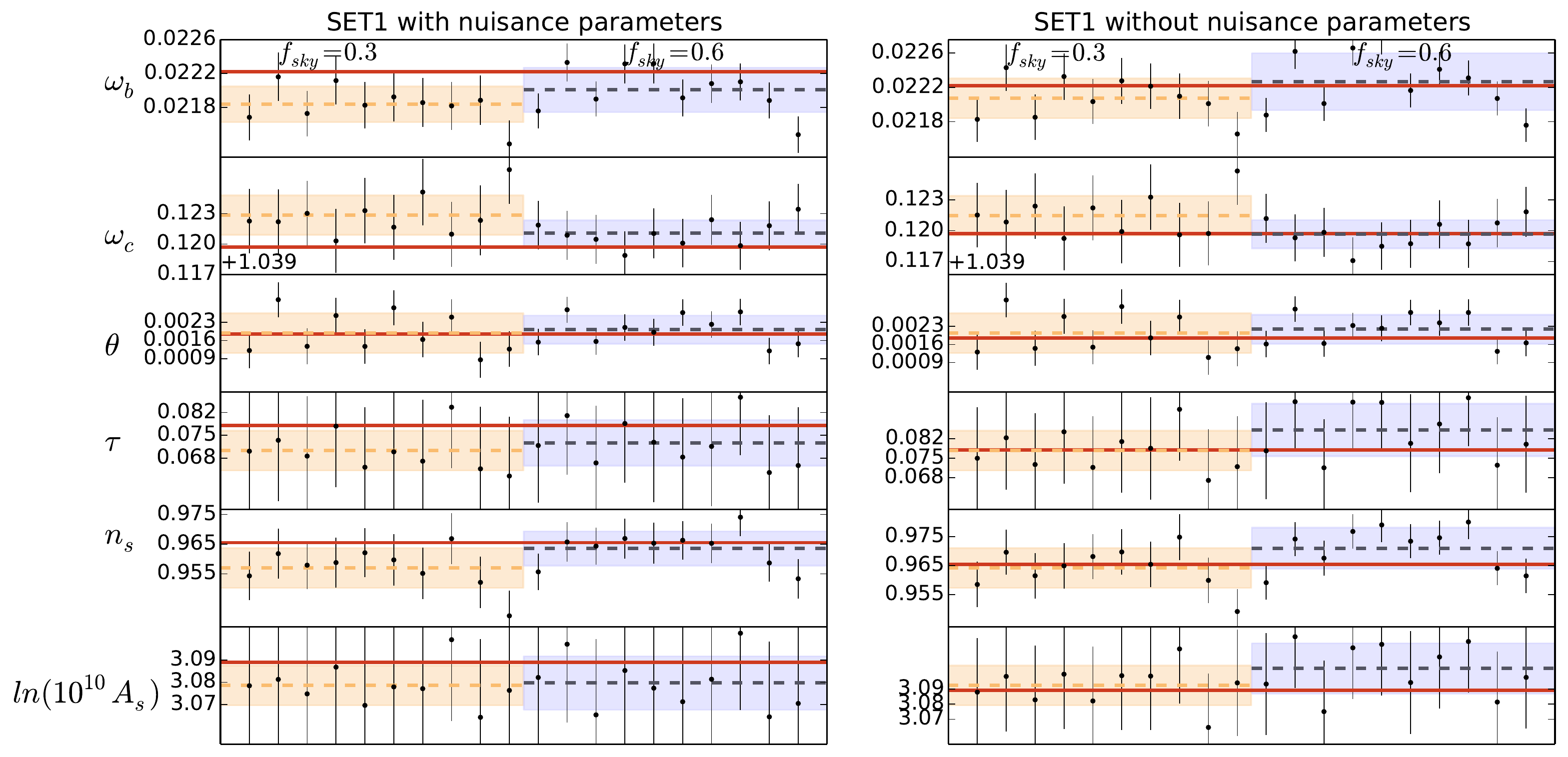}
\caption{Cosmological parameters for a subset of 10 simulation of SET1 with (left) and without (right) a model for nuisance parameters in the likelihood. Results are presented for 
$f_{sky}$=0.3 in yellow and $f_{sky}$=0.6 in blue. For each $f_{sky}$, the dashed line represents the average of the simulations marginal means and the shaded
 band represents the 1$\sigma$ scatter around this average. Each dot represents the marginal mean and 68\% CL error bar 
of the parameters in a given simulation. The red line shows the input parameters of theoretical $C_{\ell}$ used for simulations.}\label{fig_cosmo_sim}
\end{figure*}

\begin{figure*}
\centering
\includegraphics[width=\hsize]{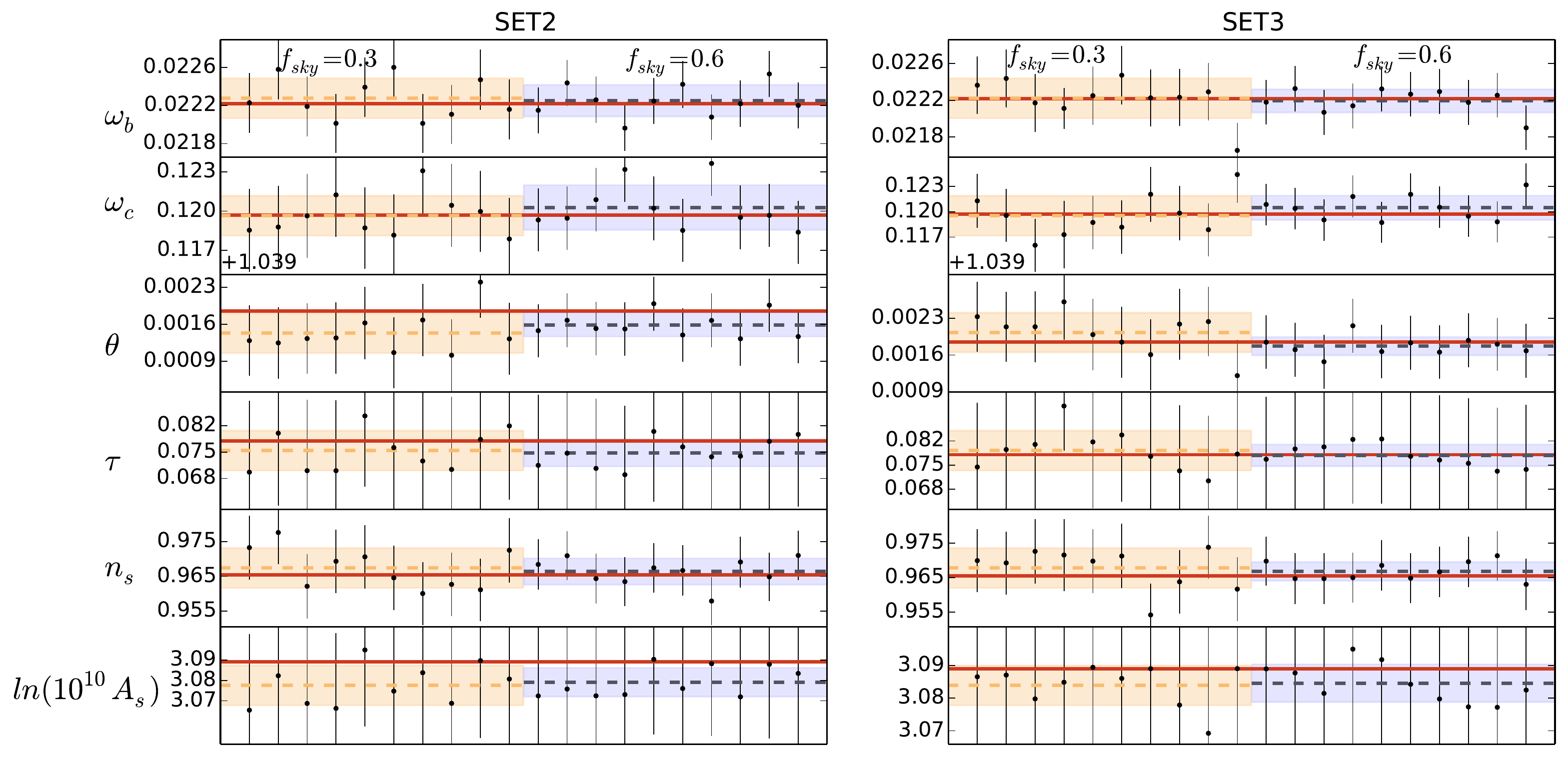}
\caption{Cosmological parameters for a subset of 10 simulation of SET2 (left) and SET3 (right). Results are presented for 
$f_{sky}$=0.3 in yellow $f_{sky}$=0.6 in blue. For each $f_{sky}$, the dashed line represents the average of the simulations marginal means and the shaded
 band represents the 1$\sigma$ scatter around this average. Each dot represents the marginal mean and 68\% CL error bar 
of the parameters in a given simulation. The red line is the input of theoretical $C_{\ell}$ used for simulations.}\label{fig_cosmo_sim2}
\end{figure*}
We test our approach by obtaining cosmological parameters from  the SMICA best fit angular power spectrum. We do this both on Planck data and on a subset of simulations. We compare the parameters obtained from simulations to the input ones. The parameters obtained from Planck data are compared to the baseline Planck 2015 results. Since we have only temperature data, we use a Gaussian prior on the parameter $\tau$: this configuration in \citet{P15_lkl} is referred to as \textit{PlikTT+tauprior}.
For each case studied, we run Monte Carlo Markov Chains (MCMC) with CosmoMC \citep{cosmomc} in combination with 
PICO\footnote{available for download at https://github.com/marius311/pypico} \citep{pico}. 
We also cross-check some of our runs using CosmoMC with CAMB \citep{camb}, 
and using CosmoSlik \citep{CosmoSlik} with PICO: we observe that results are consistent with those obtained using CosmoMC with PICO.  
For this reason, all the results presented in this analysis are obtained using the latter configuration.

\subsection{The likelihood}
 We build our likelihoods from the best fit CMB spectra obtained from the SMICA fit for the different cases under analysis. 
We use an idealised form for the likelihood, which considers no intermode correlations. This approximation should not strongly affect our results since we use bins of $\Delta \ell$ = 15. The likelihood takes the form:
\begin{equation}
-\ln \mathcal{L} \Big(\hat{C} | C(\theta)\Big)  = \frac{1}{2} \Big( \hat{C} - C(\theta) \Big)~ \Sigma^{-1} ~\Big( \hat{C} - C(\theta)\Big) +c
\end{equation}
where $\hat{C}$ and $C(\theta)$ are the best fit and theoretical angular power spectra respectively, $\Sigma$ is the covariance matrix given by the SMICA error bars on the best fit and $c$ is a constant.
The error bars are an estimate derived from the Fisher matrix. They represent the cosmic variance, foregrounds, noise and mask contribution to the error budget, but do not include uncertainties on calibration and beams.  \\ 
We explore a minimal $\Lambda$CDM model  with two approximately massless neutrinos and one massive neutrino with $\sum {m_{\nu}}=0.06$ eV.
We also use a Gaussian prior on the optical depth to reionization: for the MCMC on data we use $\tau = 0.07 \pm 0.02$, the same as in Planck analysis \citep{P15_lkl}, while for simulations
we choose $\tau = 0.078 \pm 0.02$, since $\tau$ = 0.078 corresponds to the input value of the simulated CMB maps.\\
There is a small amount of foreground residuals in the CMB spectra, as evident from Fig. \ref{fig_cmb}. 
This residual has to be accounted for in the likelihood formulation with a nuisance model. Finding a shape for the foreground residuals is not trivial, since nuisance parameters can bias the cosmological parameters when incorrectly chosen. We opt for a physical modeling of the nuisance parameters based on our foreground knowledge. \citet{Paoletti:2012} find that two terms for the shot noise and clustered contribution suffice to account for the background galaxies contribution.  
Also, we need to account for residuals of the galactic dust. We do not consider any term for the SZ residual contamination.\\
The \textit{Planck Collaboration} derives cosmological parameters from the CMB maps, including the SMICA one \citep{P15_compsep}. The SMICA map cosmological parameters cannot be directly compared to this analysis parameters since the map-making procedure
can add some foreground contribution, thus we compare our results with those obtained with the Planck likelihood, which uses angular power spectra of data maps.
Nevertheless, similarly to what it is done in the Planck analysis on CMB maps , we use a nuisance model that comprises:
\begin{itemize}
 \item a point source term with flat spectrum. Its amplitude is regulated by the parameter $A^{PS}$, which corresponds to the point sources contribution for $D_{\ell=3000}$;
 \item a clustered CIB term  with a spectrum  $\ell^{n_{CIB}}$. We fix $n_{CIB}$ = -1.3 for most explorations, unless otherwise stated. The amplitude $A^{CIB}$ represents the CIB contribution for $D_{\ell=3000}$;
 \item a dust term with an angular power spectrum $\ell^{-2.6}$. The nuisance parameter $A^{dust}$ is defined as the emission for $C_{\ell=500}$.
\end{itemize}
The physical nuisance model is our reference configuration. 
In a subset of cases we also test using a smaller number of nuisance parameters, as well as the use of a single template derived from simulations. 
The template is based on the shape of the average foreground residuals in SET3 simulations at the largest $f_{sky}$, i.e., the case with the 
strongest residual contamination in the CMB spectrum. Its shape does not represent any particular foreground contamination, however it is very close 
to the clustered CIB theoretical shape, meaning that this is the major contribution that we expect in the residuals according to simulations. 
Fig.~\ref{fig_template} shows the difference between the best fit CMB spectrum and the spectrum of the input CMB map for 10 SET3 simulations. 
The input maps are unmasked, thus the low-$\ell$ scatter is largely driven by cosmic variance. From these we compute the average residuals and 
fit a shape for the template. We use this template as a unique nuisance component in the likelihood exploration, only changing its amplitude.

\subsection{Cosmological parameters from simulations}
\begin{figure*}
\centering
\includegraphics[width=\hsize]{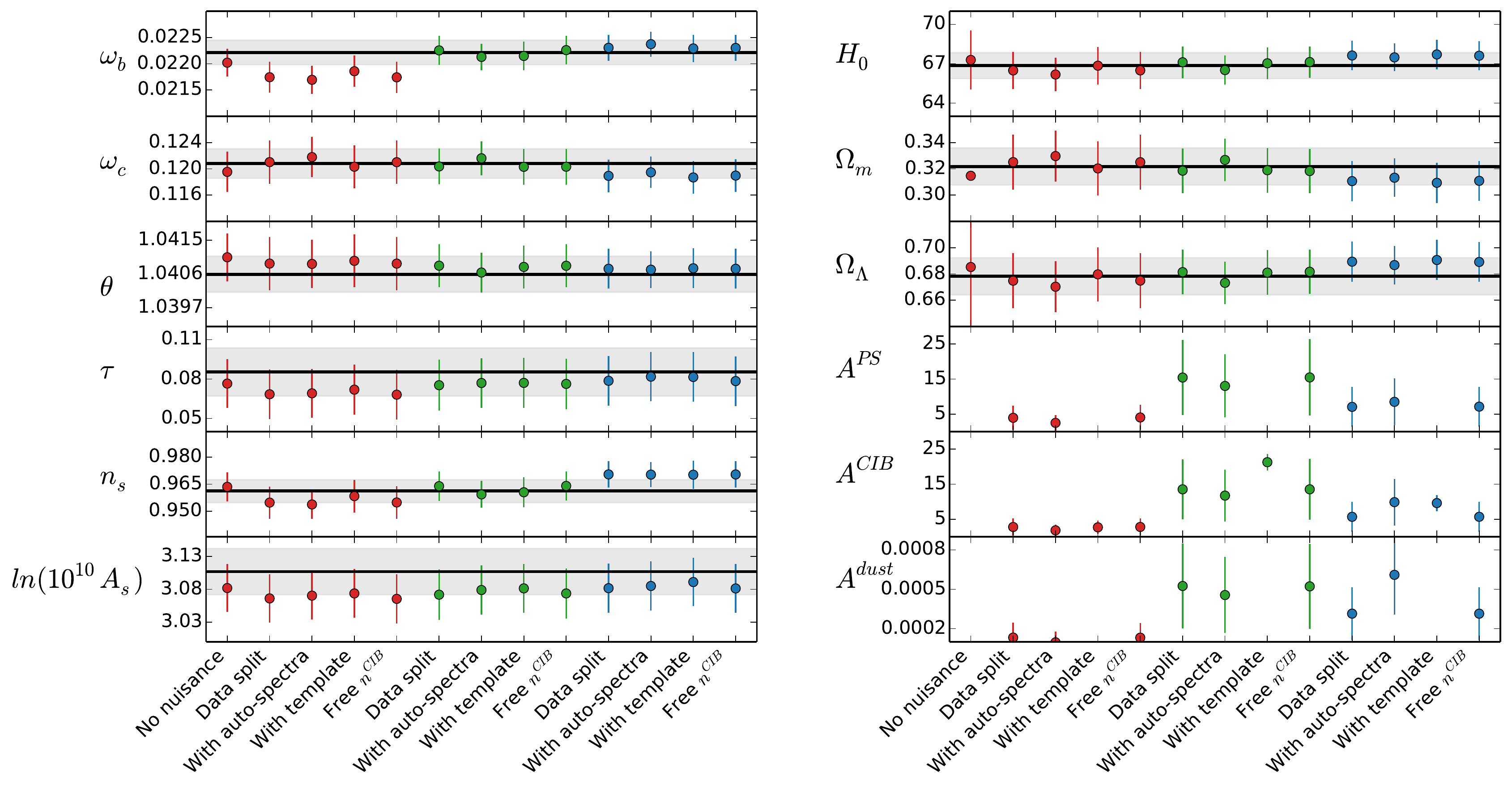}
\caption{ Marginal mean and 68\% CL error bars for cosmological parameters obtained with Like-F03 in red, Like-F05 in green and Like-F06 in blue. For comparison, 
Planck 2015 high-$\ell$ temperature likelihood results are shown in the background, with the marginal mean drawn in black and 68\% CL as a grey band. 
On the left we show the six standard $\Lambda CDM$ cosmological parameters and on the right some derived and nuisance parameters. Nuisance parameters 
show no Planck comparison since they are specific to the likelihood. The values of nuisance parameters are in units of $\mu K^2$. 
The template amplitude is shown in the $A^{CIB}$ column since its shape is very close to that of the CIB term in the nuisance physical model.}\label{fig_cosmo}
\end{figure*}
We explore cosmological parameters for the first 10 simulations of each set. For these simulations, we obtain parameters for both the largest and the smallest masks, to check for effects that depend on retained  sky fraction. For each simulation and sky fraction, we use the best fit CMB angular power spectrum of SMICA to build a likelihood as described in the previous section. The main analysis is done using the physical parameterization of the nuisance model.
The list of all parameters is detailed in the first column of Table \ref{tab1}.
Cosmological parameters are presented in Fig.~\ref{fig_cosmo_sim} for SET1 and Fig.~\ref{fig_cosmo_sim2} for SET2 and SET3, where the 
red line shows simulation inputs and the wide coloured band show 1$\sigma$ scatter of the marginal mean values.\\
SET1 simulations are those that best recover the input CMB power spectrum in the SMICA fit, thus we expect their residual foreground 
content to be very low. 
We test SET1 simulations  in two different configurations, letting the nuisance parameters free and setting all of them to zero, i.e., 
not accounting for any residuals in the likelihood exploration. As shown in the right panel of Fig.~\ref{fig_cosmo_sim}, 
the MCMC exploration with nuisance parameters shows evident biases with both masks. As a cross-test, we obtain cosmological parameters from theoretical spectra to which we add some scatter according to the expected cosmic variance. In this case the average parameters obtained coincide with the input, meaning that the shift we observe in Fig.~\ref{fig_cosmo_sim} are due to foreground residuals and not to our pipeline implementation. \\
As we will see in Sect.~\ref{cosmo_param_data}, there is a degeneracy between the shape of the foreground residuals and the cosmological parameters, and a wrong estimation of the nuisance parameters can induce biases. In particular, when we have a low $f_{sky}$, the error bars are larger and we can more easily mix up the CMB and the foregrounds. This is evident in Fig.~\ref{fig_cosmo_sim}, which shows how most biases are strongly reduced when nuisance parameters are removed,  especially for $f_{sky}$ = 0.3, where we expect to have the lowest, and thus most degenerate, residuals.
Due to the low level of residuals in the CMB spectra, the nuisance parameters are not well estimated and are in most cases compatible with zero. A level of residuals is present in the data, but since this is not well determined, it is not correctly accounted for.\\
For SET2 and SET3 simulations we obtain  less biased results: in Fig. \ref{fig_cosmo_sim2} we can see that biases of parameters are 
less evident, especially for the SET3 case. In this case we did not run the nuisance-free likelihood given that the level of foreground 
residuals is too high to justify such a test. Since the level of residuals in these simulations is higher than in SET1, it is better constrained in the parameters exploration.
We observe very small changes with sky fraction, the most relevant one being the decrease in size of the 1$\sigma$ scatter band with increasing  $f_{sky}$, as expected.
We note that marginal errors on $\tau$ and $A_s$ for individual simulations are quite large, while the scatter of the mean is not: this is not surprising since the marginal error on $\tau$, and consequently on $A_s$,  is regulated by the Gaussian prior $\tau$ = 0.07 $\pm$ 0.02 we impose.\\
Since the nuisance parameters are not well constrained, a model for the residuals with less parameters could in principle reduce the 
uncertainty in the exploration.
For the SET3 case only, we test the template configuration of the likelihood described in the previous section. 
In this configuration only one nuisance parameters is fitted, which is the amplitude of the template. 
In terms of biases, the results are equivalent to those obtained with the physical nuisance model. 
The only relevant change is that the discrepancy on $A_s$ is reduced, while that on $\omega_b$ is increased. 
This suggests that the average foreground contamination represented by the template does not fully describe the details of the 
residuals in each CMB spectrum of simulations, and that the details of foreground modeling in the likelihood are important for 
accurate estimation of cosmological parameters.

\begin{table*}[t!]
\caption{\label{tab1} 
Results of the MCMC exploration with the three considered likelihoods: Like-F03, Like-F05 and Like-F06. 
Constraints on parameters are given as
marginal mean with 68\% CL error on: main parameters (top), nuisance parameters (middle) and derived parameters (bottom). 
The Planck quoted values are obtained with Planck 2015 high-$\ell$ likelihood.}
\centering

\begin{tabular}{|l|ccc|}
\hline
\hline
Parameter                       & Like-F03               &  Like-F05              &  Like-F06          \\
\hline
$ \omega_\mathrm{b}h^2$              & $0.02174 \pm 0.00030$  & $ 0.02225 \pm 0.00027  $ & $ 0.02230 \pm 0.00025 $ \\
$ \omega_\mathrm{c}h^2$             & $0.1210  \pm 0.0033$   & $ 0.1204  \pm 0.0027   $ & $ 0.1189  \pm 0.0025  $\\
$100\theta_{MC}$                    &$1.04088 \pm 0.00071$  & $ 1.04082 \pm 0.00057  $ & $ 1.04074 \pm 0.00053 $ \\
$\ln \left(10^{10}A_\mathrm{s}\right)$     &  $3.066   \pm 0.037$    & $ 3.072   \pm 0.039    $ & $ 3.082   \pm 0.037   $ \\
$n_{\mathrm s}$                &$0.9548  \pm 0.0089$   & $ 0.9639  \pm 0.0081   $ & $ 0.9704  \pm 0.0073  $ \\
$\tau$                         &$0.068   \pm 0.019$    & $ 0.075   \pm 0.019    $ & $ 0.079   \pm 0.019   $ \\
                               &                      &                          &                 \\
\hline
$A^{PS}$                       &  $3.9     \pm 3.4$      & $ 15      \pm 11       $ & $ 7.0     \pm 5.6     $ \\
$A^{CIB}$                      & $2.8     \pm 2.4 $     & $ 1341    \pm 8.6      $ & $ 5.7     \pm 4.3     $ \\
$A^{dust}$                     &  $0.00013 \pm 0.00011 $ & $ 0.0052 \pm 0.00032  $ & $ 0.00031 \pm 0.00020 $\\
\hline
$H_0$                          &  $66.5    \pm 1.4$      & $67.1     \pm 1.2      $ & $ 67.6    \pm 1.1     $ \\
$\Omega_\mathrm{m}$            &   $0.325   \pm 0.021$    & $0.318    \pm 0.017    $ & $ 0.311   \pm 0.015   $ \\
$\Omega_\mathrm{\Lambda}$      &  $0.675   \pm 0.021$    & $0.681    \pm 0.017    $ & $ 0.690   \pm 0.015   $ \\
\hline 
\end{tabular}

\end{table*}
\subsection{Cosmological parameters from Planck data}\label{cosmo_param_data}

\begin{figure*}
\centering
\includegraphics[width=\hsize]{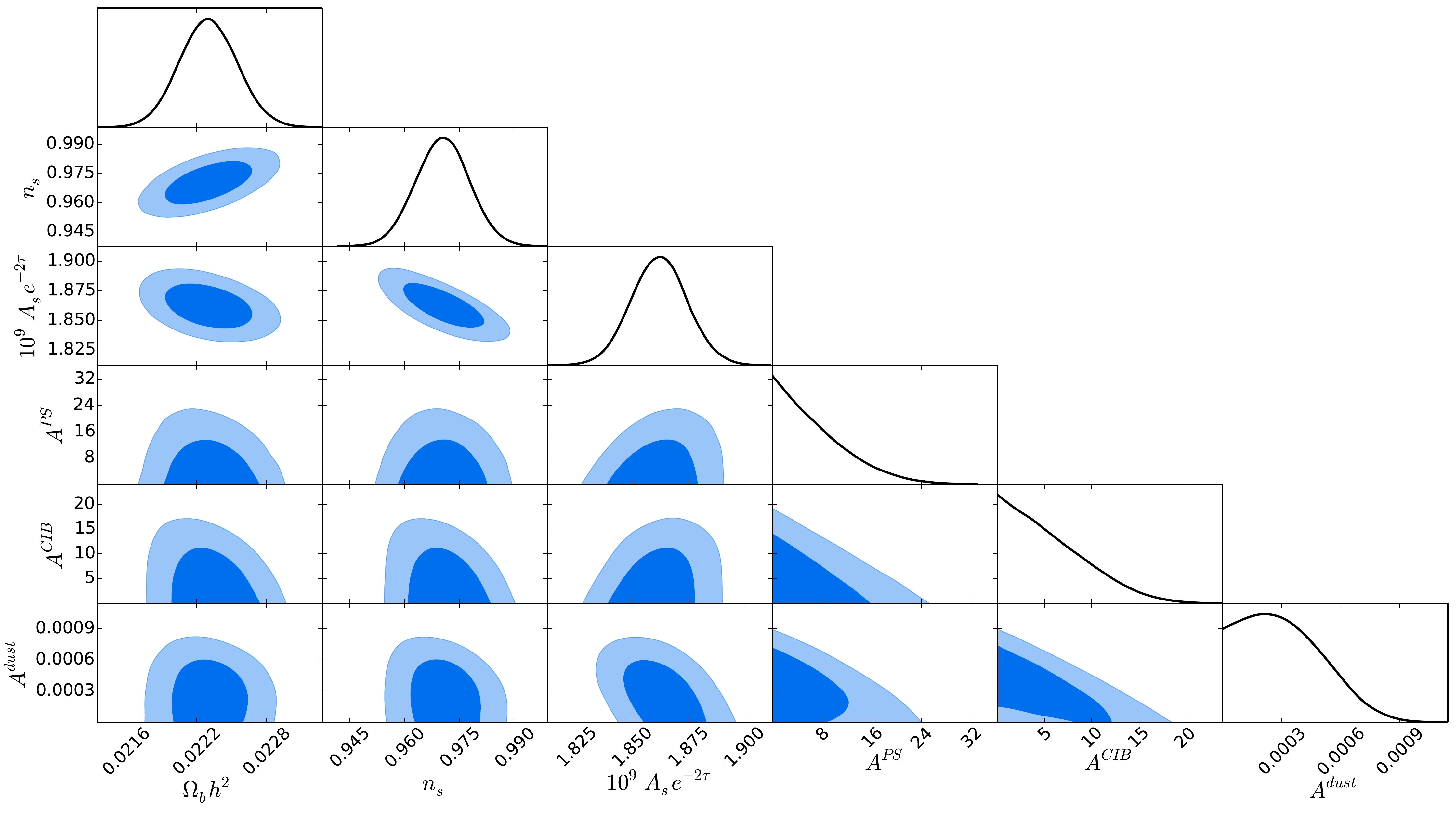}
\caption{Triangle plot showing the relation between the main cosmological parameters and the nuisance parameters, for the analysis on data with $f_{sky}$ = 0.6. Similar plots are obtained for SET3 simulations and for different $f_{sky}$. The blue and light-blue contours represent the 68\% and 95\% CL respectively.}\label{tri}
\end{figure*}

We build a likelihood for each mask from best fit spectra obtained from the analysis detailed in Sect. \ref{sec:data_analis}.
We call these three likelihoods Like-F03, Like-F05 and Like-F06, where FX refers to the $f_{sky}$ of the mask used. 
We run a MCMC exploration with Planck high-$\ell$ temperature likelihood and compare the results with the Planck published ones. We find good agreement between these two run of the Planck likelihood,  meaning that our configuration is the same as that used for the Planck analysis.\\
We give results for our three likelihoods and compare them to the Planck likelihood run.
We note that a Gaussian prior is imposed on the absolute map calibration for Planck likelihood $y_{cal} = 1 \pm 0.0025$, while
we keep this value fixed to $y_{cal}$ = 1.
We adopt this choice after testing that including this parameter in the explorations does not affect the results, 
apart from increasing the total number of parameters to be sampled.\\
We plot a comparison of the cosmological parameters estimated in Fig.~\ref{fig_cosmo},  
while the full list including derived parameters can be found in Table~\ref{tab1}.
The respective values for the Planck likelihood analysis can be found in \citet{P15_lkl}, where Table 17 lists the cosmological parameters and Table 10 the nuisance parameters.
Shifts of cosmological parameters 
in units of 1$\sigma$ Planck error bars are presented in Table \ref{tab2}: for most parameters we observe a progressive shift increasing with 
the retained sky fraction, the most evident case being for $n_s$. On the whole, parameters show at most 1$\sigma$ deviation with respect to 
the Planck analysis, with the only exception of $\omega_b$ for $f_{sky}$ = 0.3, which shows a deviation of -2.04$\sigma$, and $n_s$ for $f_{sky}$ = 0.6, which shows a deviation of 1.48$\sigma$.\\
At low $f_{sky}$ the residuals are weak and not clearly constrained by the nuisance model. As seen in simulations, there is an uncertainty in the value of the nuisance parameters that induces a shift in the cosmological parameters. In Fig.~\ref{tri} we can see that  nuisance parameters are consistent with zero and are also strongly degenerate among them. Some degeneracies are also visible with the cosmological parameters, as for example between $A_s e^{-2\tau}$ and $A^{dust}$. 
Due to the strong correlation between all cosmological parameters, these degeneracies can induce biases. 
Also, as noted by \citet{Huffenberger:2006}, the parameter $n_s$ is particularly sensitive to incorrect subtraction of the point source 
component,  since a residual of point sources can mimic a different tilt of the CMB angular power spectrum. 
The biases we obtain are representative of the uncertainty on the determination of the foreground model.  \\
For Like-F03 we perform an analysis without any nuisance parameters, finding smaller shifts with respect to the reference configuration. 
This is further evidence for the existence of degeneracies between cosmological and weakly constrained nuisance parameters. 
Results are shown in the ``No nuisance'' column of Fig.~\ref{fig_cosmo}. This analysis is only possible for $f_{sky}$=0.3 since the level of residuals is too high 
for the other two masks. \\
The results of the analysis using a template are also shown in Fig. \ref{fig_cosmo}.  
This configuration performs slightly better, especially at low $f_{sky}$, reducing the biases on $n_s$ and $\omega_b$. 
No significant improvement is seen however at high  $f_{sky}$. The more, the total foreground power detected by the template is 
lower at $f_{sky}$ = 0.6 than at $f_{sky}$ = 0.5. This suggests that for high sky fraction the template is not anymore 
representative  of the residual contamination in the CMB spectrum. This is partially true also for the physical nuisance model, 
meaning that the residuals at large sky fraction are not well represented by neither model.\\
We note that the biases observed in the data analysis are different from those observed in simulations. This suggests that the foreground complexity is not well accounted for in our simulations and that the  correct choice of the nuisance model strongly depends on the details of the foreground contamination. 

\subsection{Cross-tests on data}
Results presented in Tables \ref{tab2} and \ref{tab1} refer to the main exploration detailed above. As a cross-check we present in Fig. \ref{fig_cosmo} results of
two different configurations.
The first uses best fit CMB spectra as obtained from Sect. \ref{sec:nosplit}, that is using both auto- and cross-spectra to build
covariance matrices.
The second adds as nuisance parameter the CIB index which defines the angular power spectrum shape of the CIB residuals as $\ell^{n^{CIB}}$.
This parameter is varied with a Gaussian prior $n^{CIB} = -1.3 \pm 0.2$. 
We observe no relevant shift in the obtained cosmological parameters from this two additional configurations. \\
We also obtain cosmological parameters from the best fit obtained using the 857 GHz channel as described in Sect. \ref{sec:857}. 
We run MCMC for $f_{sky}$ = 0.6 on SET3 simulations and obtain cosmological parameters which are consistent with those shown in Fig. \ref{fig_cosmo_sim2} within maximum 0.022$\sigma$ ($\sigma$ here is the scatter of the marginal mean among various simulations).
Instead, on the Planck data analysis, cosmological parameters are more strongly biased than those obtained without the 857 GHz channel, in particular $n_s$ and $\omega_b$. While the increase in foreground residuals in the spectrum is modest, we expect  their characteristics to be quite complex and not adjustable by the minimal nuisance model we use.

\begin{table}
\caption{\label{tab2} 
Shift of parameters between the three data likelihoods Like-F03, Like-F05 and Like-F06 and the Planck high-$\ell$ likelihood results 
in units of 1$\sigma$ 
Planck errors.}
\centering
\begin{tabular}{|l|ccc|}
\hline
\hline
 Parameter                     & Like-F03    &  Like-F05 &  Like-F06 \\
\hline
$ \omega_\mathrm{b}h^2$        &  $ -2.04  $ & $  0.17 $ & $  0.39 $ \\
$ \omega_\mathrm{c}h^2$        &  $  0.11 $  & $ -0.19 $ & $ -0.84 $ \\ 
$100\theta_{MC}$               &  $  0.59  $ & $  0.48 $ & $  0.30 $ \\ 
$\tau$                         &  $ -0.93  $ & $ -0.55 $ & $ -0.37 $ \\ 
$n_{\mathrm s}$                &  $ -1.02  $ & $  0.43 $ & $  1.48 $ \\ 
$\log \left(10^{10}A_\mathrm{s}\right)$&$-1.17$& $-1.00$ & $ -0.72 $ \\ 
$H_0$                          &  $ -0.37  $ & $  0.26 $ & $  0.79 $ \\ 
$\Omega_\mathrm{m}$            &  $  0.24  $ & $ -0.23 $ & $ -0.79 $ \\ 
$\Omega_\mathrm{\Lambda}$      &  $ -0.24  $ & $  0.23 $ & $  0.79 $ \\

\hline
\end{tabular}
\end{table}

\section{Conclusions}\label{sec:conclusion}

We have studied a new configuration of the SMICA method to estimate the CMB angular power spectrum directly via component separation.
This configuration uses only cross-angular power spectra between half-mission data split, thus avoiding the noise bias present in autospectra.
We use a constrained foreground model that targets the extragalactic point sources emission. This is particularly important since the point sources level is degenerate with the CMB at small scales. 
Using SMICA, we jointly fit for the CMB power spectrum, the point sources emission law and other foregrounds angular spectra and frequency emission, such as dust and clustered CIB.\\ 
We obtain an estimate for the point sources emission law that is consistent with independent estimates in \textit{Planck Collaboration} analysis.
We recover a fit of the CMB angular power spectrum that we use to derive cosmological parameters through an MCMC 
likelihood exploration, both on Planck 2015 data and simulations. To model the foreground residuals in the CMB spectra, two configurations of nuisance parameters are studied, a physical model and an artificial template based on results of simulations.
In both cases the cosmological parameters we obtain for simulations and Planck data agree with the predicted values of the \textit{Planck Collaboration} 
analysis \citep{P15_cosmopar} within  1$\sigma$ on average. The level of biases that we observe on the simulations shows us the level 
of bias to expect on the real data. \\
The observed shifts strongly depend on the foreground residuals and on the nuisance model of the likelihood. 
If the foreground residuals are weak, the nuisance parameters are not well constrained by the data, and their misevaluation can induce biases on the cosmological parameters. When the foreground contamination is stronger the foreground model is better constrained and the biases are reduced. 
However the residuals characteristics need careful modeling, and a minimal model as that used in this work is not sufficient to describe them.
We observe this both on simulations and on Planck data.
Using a single foreground template for nuisance has the advantage of having less parameters to fit, however the shape of this template is not universal and again depends on the foregrounds characteristics. \\
In conclusion we observe that a blind method as SMICA with an adapted model for the extragalactic foregrounds can recover an 
estimate of the CMB power spectrum which has a very low foreground residual content. However, in order to use this estimate for 
cosmological purposes, a deep knowledge and careful modeling of the shape of the foreground residuals is needed. The same is true also for the Planck likelihood analysis in \citet{P13_lkl}, thus it is not clear whether this blind approach grants any advantage, at least for the CMB temperature analysis, where foregrounds are particularly complex. When considering polarization, a blind approach could still be useful since the small scales foreground contamination is more simple, while SMICA has demonstrated its ability to deal with large scale complex contamination \citep{P13_compsep}.
Also, using a zone approach with SMICA that would take into account the variability of galactic foregrounds on the sky could improve the efficiency of component separation and thus improve the cosmological parameters estimation.

\begin{acknowledgements}
UC wishes to thank Aur\'elien Benoit-L\'evy for suggestions on masks treatment, Silvia Galli for discussions on simulations and cosmological parameters,
\'Eric Hivon for tips on the use of HEALPix,
Marius Millea for helpful insight on radio point sources and CosmoSlik and Mario Ballardini for helpful suggestions on CosmoMC.
This work has made use of the Horizon Cluster hosted by Institut d'Astrophysique de Paris.
We thank Stephane Rouberol for running smoothly this cluster for us.
UC has been supported within the Labex ILP (reference ANR-10-LABX-63) part of the Idex SUPER, and received financial state aid
managed by the Agence Nationale de la Recherche, as part of the programme Investissements
d’avenir under the reference ANR-11-IDEX-0004-02.
This research used publicly available Planck data.
\end{acknowledgements}

\bibliographystyle{aa} 
\bibliography{bib_smica_paper} 
\begin{appendix}

\section{Masks}\label{appa}
The masks we use are the sum of a galactic and a point source part: while the point source part is the same for all masks, the galactic sky coverage changes.
For the galactic part we create a set of masks starting from those delivered by the {\it Planck Collaboration}\footnote{available 
for download at http://pla.esac.esa.int ({\it HFI\_Mask\_GalPlane-apo0\_2048\_R2.00.fits})}. 
From these, we choose the three galactic masks of retained sky fraction $f_{sky}$ = 0.4, 0.6, 0.7.
These mask are not apodised, thus need to be smoothed at the edges.
First we smooth them with a Gaussian beam of $FWHM=3^o$ and then we threshold them 
to obtain a new set of slightly smaller masks of $f_{sky}$ = 0.45, 0.65, 0.75. 
This step is needed in order to avoid that the subsequent apodisation results in a large decrease of the retained sky fraction. 
Using the {\it process\_mask} function of the HEALPix package, for each of these masks we obtain a distance map, i.e., 
a map in which each pixel contains the distance to the nearest masked pixel. Such a map is used to apodise the galactic masks by convolution with 
a Gaussian of $FWHM= 4^o$.
The use of distance maps instead of a simple Gaussian smoothing avoids 
leakage into the original mask. \\
We create the point source mask based on the Planck 2015 Catalogue of Compact Sources\footnote{v2.0, also available for download at http://pla.esac.esa.int} as the union of the point sources masks at the five frequency channels 
of interest.
This point source mask is apodised with $FWHM=1^o$, using a distance map. 
We combine our point source apodised mask with our apodised galactic masks to obtain the final set of masks we use in this analysis.
Their respective retained sky fraction is $f_{sky}$=0.31, 0.48, 0.57, but to enhance readability we refer to them as
$f_{sky}$=0.3, 0.5, 0.6 throughout the text. \\
We also create a second set of masks for cross-check analysis that include point sources at 857 GHz. These masks have a similar sky fraction
to the leading set, and they are used only for the tests performed in Sect. \ref{sec:857} \\

\end{appendix}

\end{document}